\documentclass[a4paper,12pt]{iopart}
\usepackage{iopams}
\usepackage{setstack}
\usepackage{graphicx}
\usepackage{bm}
\usepackage{epsfig}
\newcommand{\diag}{{\rm diag\,}}
\newcommand{\Str}{{\rm Str\,}}
\newcommand{\Sdet}{{\rm Sdet\,}}
\newcommand{\UOSp}{{\rm UOSp\,}}
\newcommand{\U}{{\rm U\,}}
\newcommand{\Herm}{{\rm Herm\,}}

\newcommand{\eins}{\leavevmode\hbox{\small1\kern-3.8pt\normalsize1}}
\eqnobysec

\begin{document}

\newtheorem{definition}{Definition}[section]
\newtheorem{assumption}[definition]{Assumption}
\newtheorem{theorem}[definition]{Statement}
\newtheorem{lemma}[definition]{Lemma}
\newtheorem{corollary}[definition]{Statement}

\title[Random matrix ensembles and supersymmetry]{Arbitrary rotation invariant random matrix ensembles and supersymmetry:
orthogonal and unitary--symplectic case}
\author{Mario Kieburg$^{a)\dagger}$, Johan Gr\"onqvist$^{b)}$ and Thomas Guhr$^{a)}$}
\address{\ $^{a)}$Universit\"at Duisburg-Essen, Lotharstra\ss e 1, 47048 Duisburg, Germany\\
\ $^{b)}$Matematisk Fysik, LTH, Lunds Universitet, Box 118, 22100 Lund, Sweden}
\eads{$^\dagger$ \mailto{mario.kieburg@uni-due.de}}
\date{\today}

\begin{abstract}
  Recently, the supersymmetry method was extended from Gaussian
  ensembles to arbitrary unitarily invariant matrix ensembles by
  generalizing the Hubbard--Stratonovich transformation. Here, we
  complete this extension by including arbitrary orthogonally and
  unitary--symplectically invariant matrix ensembles. The results are
  equivalent to, but the approach is different from the
  superbosonization formula. We express our results in a unifying way.
  We also give explicit expressions for all one--point functions and
  discuss features of the higher order correlations.
\end{abstract}

\pacs{02.30.Px, 05.30.Ch, 05.30.-d, 05.45.Mt}
\submitto{\JPA}
\maketitle

\section{Introduction}\label{sec1}
In random matrix theory, supersymmetry is an indispensable tool
\cite{Efe83,VWZ85,Efe97,GMW98}. Recently, this method was extended
from Gaussian probability densities to arbitrary rotation invariant
ones. Presently, there are two approaches referred as
superbosonization. The first approach is a generalization of the
Hubbard--Stratonovich transformation for rotation invariant random
matrix ensembles \cite{Guh06}. The basic idea is the introduction of a
proper Dirac--distribution in superspace, extending earlier work in
the context of scattering theory \cite{LSSS95}, universality
considerations \cite{HacWei95}, field theory \cite{EST04,EfeKog04} and
quantum chromodynamics \cite{BasAke07}. The second approach is the
superbosonization formula developed in Refs.  \cite{Som07,LSZ07}. It
is an identity for integrals over superfunctions on rectangular
supermatrices which are rotation invariant under an ordinary group.

Here, we further extend the generalized Hubbard--Stratonovich
transformation to the orthogonal and the unitary symplectic symmetry
class in a unifying way. To this end, we use an analog of the
Sekiguchi differential operator for ordinary matrix Bessel--functions.
We also aim at a presentation which is mathematically more sound than
the one in Ref.  \cite{Guh06}.

The article is organized as follows. The problem is posed in Sec.
\ref{sec2}. We give an outline of the calculation in Sec.
\ref{sec2.5}. In Sec. \ref{sec3}, we present the generalized
Hubbard--Stratonovich transformation. In Sec. \ref{sec4}, we carry out
the calculation for arbitrary ensembles as far as possible. Then, we
restrict the computation to the three classical symmetry classes. We,
thereby, extend the supersymmetric Ingham--Siegel integral
\cite{Guh06}. In Sec. \ref{sec5}, we give a more compact expression of
the generating function in terms of supermatrix Bessel--functions. We
show that the generating function is independent of the chosen
representation for the characteristic function. The one--point and
higher correlation functions are expressed as eigenvalue integrals in
Sec. \ref{sec6}. In the appendices, we present details of the
calculations.

\section{Posing the problem}\label{sec2}

We consider a sub-vector space $\mathfrak{M}_N$ of the hermitian
$N\times N$--matrices $\Herm(2,N)$. $\Herm(\beta,N)$ is the set of
real orthogonal ($\beta=1$), hermitian ($\beta=2$) and quaternionic
self-adjoint ($\beta=4$) matrices and $\beta$ is the Dyson-index. We
use the complex $2\times2$ dimensional matrix representation for
quaternionic numbers $\mathbb{H}$. The results can easily be extended
to other representations of the quaternionic field. For the relation
between the single representations, we refer to a work by Jiang
\cite{Jia05}.

The object of interest is an arbitrary sufficiently integrable
probability density $P$ on $\mathfrak{M}_N$. Later, we assume that $P$
is an invariant function under the action of the group
\begin{equation}\label{2.1}
 \U^{(\beta)}(N)=\left\{\begin{array}{ll}
             {\rm O}(N)	&	,\ \beta=1\\
	     \U(N)	&	,\ \beta=2\\
	     {\rm USp}(2N)	&	,\ \beta=4
            \end{array}\right.
\end{equation}
and $\mathfrak{M}_{\gamma_2N}=\Herm(\beta,N)$. Here, we introduce
$\gamma_2=1$ for $\beta\in\{1,2\}$ and $\gamma_2=2$ for $\beta=4$ and,
furthermore, $\gamma_1=2\gamma_2/\beta$ and
$\tilde{\gamma}=\gamma_1\gamma_2$. These constants will play an
important role in the sequel.

We are interested in the $k$--point correlation functions
\begin{equation}\label{2.2}
 R_k(x)=\mathbf{d}^k\int\limits_{\mathfrak{M}_N}P(H)\prod\limits_{p=1}^k\tr\delta(x_p\eins_N-H)d[H]
\end{equation}
with the $k$ energies $x=\diag(x_1,\ldots,x_k)$. Here, $\mathbf{d}$ is
the inverse averaged eigenvalue degeneracy of an arbitrary matrix
$H\in \mathfrak{M}_N$. The measure $d[H]$ is defined as in Ref.
\cite{KKG08}, it is the product of all real and imaginary parts of the
matrix entries. For example, we have $\mathbf{d}=1/2$ for
$\mathfrak{M}_{2N}=\Herm(4,N)$ and $\mathbf{d}=1$ for no eigenvalue
degeneracy as for $\mathfrak{M}_{N}=\Herm(\beta,N)$ with
$\beta\in\{1,2\}$. We use in Eq. \eref{2.2} the $\delta$--distribution
which is defined by the matrix Green's function. The definition of the
$k$--point correlation function \eref{2.2} differs from Mehta's
\cite{Meh67}. The two definitions can always be mapped onto each other
as explained for example in Ref.  \cite{GMW98}.

We recall that it is convenient to consider the more general function
\begin{equation}\label{2.4}
 \widehat{R}_k\left(x^{(L)}\right)=\mathbf{d}^k\int\limits_{\mathfrak{M}_N}P(H)\prod\limits_{p=1}^k\tr[(x_p+L_p\imath\varepsilon)\eins_N-H]^{-1}d[H]
\end{equation}
where we have suppressed the normalization constant. The quantities
$L_j$ in $x^{(L)}=\diag(x_1+
L_1\imath\varepsilon,\ldots,x_k+L_k\imath\varepsilon)$ are elements in
$\{\pm 1\}$. We define
$x^{\pm}=\diag(x_1\pm\imath\varepsilon,\ldots,x_k\pm\imath\varepsilon)$.
Considering the Fourier transformation of \eref{2.2} we have
\begin{eqnarray}
 r_k(t) & = & (2\pi)^{-k/2}\int\limits_{\mathbb{R}^k}R_k(x)\prod\limits_{p=1}^k\exp\left(\imath x_pt_p\right)d[x]=\nonumber\\
 & = & \left(\frac{\mathbf{d}}{\sqrt{2\pi}}\right)^k\int\limits_{\mathfrak{M}_N}P(H)\prod\limits_{p=1}^k\tr \exp\left(\imath Ht_p\right)d[H]\ .\label{2.5}
\end{eqnarray}
The Fourier transformation of \eref{2.4} yields
\begin{eqnarray}
 \widehat{r}_k(t)& = &(2\pi)^{-k/2}\int\limits_{\mathbb{R}^k}\widehat{R}_k\left(x^{(L)}\right)\prod\limits_{p=1}^k\exp\left(\imath x_pt_p\right)d[x]=\nonumber\\
 & = & \prod\limits_{p=1}^k\left[-L_p\ 2 \pi\imath\Theta(-L_pt_p)\exp\left(\varepsilon L_pt_p\right)\right] r_k(t)\label{2.6}
\end{eqnarray}
where $\Theta$ is the Heavyside--distribution.

As in Ref.
\cite{Guh06}, the $k$--point correlation function is completely
determined by Eq. \eref{2.4} with $L_p=-1$ for all $p$ if the Fourier
transform \eref{2.5} is entire in all entries, i.e. analytic in all
entries with infinite radius of convergence. We obtain such a Fourier transform if the $k$--point correlation function $R_k$ is a Schwartz--function on $\mathbb{R}^k$ with the property
\begin{equation}\label{2.9}
 \int\limits_{\mathbb{R}^k}|R_k(x)|\prod\limits_{p=1}^k\exp\left(\tilde{\delta} x_p\right)d[x]<\infty\quad,\quad\forall\tilde{\delta}\in\mathbb{R}\ .
\end{equation}
This set of functions is dense in the set of Schwartz--functions on $\mathbb{R}^k$ without this property. The notion dense refers to uniform convergence. This is true since every Schwartz--function times a Gaussian distribution $\exp\left(-\epsilon\sum\limits_{p=1}^kx_p^2\right)$, $\epsilon>0$, is a Schwartz--function and fulfils Eq. \eref{2.9}. We proof that $r_k$, see Eq. \eref{2.5}, is indeed entire in all entries for such $k$--point correlation functions. To this end, we consider the function
\begin{equation}\label{2.10}
 r_{k\delta}(t)=\int\limits_{\mathfrak{B}_\delta}R_k(x)\prod\limits_{p=1}^k\exp\left(\imath x_pt_p\right)d[x],
\end{equation}
where $\mathfrak{B}_\delta$ is the closed $k$-dimensional real ball with radius $\delta\in\mathbb{R}^+$. Due to the Paley--Wiener theorem \cite{Hoe76}, $r_{k\delta}$ is for all $\delta\in\mathbb{R}^+$ entire analytic. Let $\mathfrak{B}_{\tilde{\delta}}^{\mathbb{C}}$ be another $k$-dimensional complex ball with radius $\tilde{\delta}\in\mathbb{R}^+$. Then, we have
\begin{equation}
 \underset{\delta\to\infty}{\lim}\underset{t\in\mathfrak{B}_{\tilde{\delta}}^{\mathbb{C}}}{\sup}|r_{k\delta}(t)-r_k(t)| \leq \underset{\delta\to\infty}{\lim}\int\limits_{\mathbb{R}^k\setminus\mathfrak{B}_\delta}|R_k(x)|\prod\limits_{p=1}^k\exp\left(\tilde{\delta} x_p\right)d[x]=0\ .\label{2.11}
\end{equation}
The limit of $r_{k\delta}$ to $r_k$ is uniform on every compact support on $\mathbb{C}^k$. Thus, $r_k$ is entire analytic.

The modified correlation
function $\widehat{R}_k$ for all choices of the $L_p$ can be
reconstructed by Eq. \eref{2.6}. In Sec. \ref{sec6}, we extend the
results by a limit--value--process in a local convex way to
non-analytic functions.

We derive $\widehat{R}_k\left(x^-\right)$ from the generating function
\begin{equation}\label{2.8}
 Z_k\left(x^{-}+J\right)=\int\limits_{\mathfrak{M}_N}P(H)\prod\limits_{p=1}^k\frac{\det[H-(x_p^-+J_p)\eins_N]}{\det[H-(x_p^--J_p)\eins_N]}d[H]
\end{equation}
by differentiation with respect to the source variables \cite{Zir06}
\begin{equation}\label{2.7}
 \widehat{R}_k\left(x^{-}\right)=\left(\frac{\mathbf{d}}{2}\right)^k\left.\frac{\partial^k}{\prod_{p=1}^k\partial J_p}Z_k\left(x^{-}+J\right)\right|_{J=0}
\end{equation}
where $x^{-}+J=x^{-}\otimes\eins_4+\diag(J_1,\ldots,J_k)\otimes\diag(-\eins_2,\eins_2)$. By definition, $Z_k$ is normalized to unity at $J=0$.

\section{Sketch of our approach}\label{sec2.5}

To provide a guideline through the detailed presentation to follow in the ensuing Sections,
we briefly sketch the main ideas as in Ref.~\cite{Guh06} and as further extended in the
present contribution.

To express the generating function \eref{2.8} as an integral in 
superspace, we write the determinants as Gaussian integrals over
vectors of ordinary and Grassmann variables. We then perform the ensemble
average which is equivalent to calculating the characteristic function
\begin{equation}\label{2.5.1}
 \Phi(K)=\int P(H)\exp(\imath\tr HK)d[H]
\end{equation}
of the probability density.  The rotation invariance of $P(H)$ carries
over to $\Phi(K)$. The ordinary matrix $K$ contains the abovementioned
vectors of ordinary and Grassmann variables as dyadic matrices. It has
a dual matrix $B$ in superspace whose entries are all scalarproducts
of these vectors. The reduction in the degrees of freedom is fully
encoded in this duality, as the dimensions of $K$ and $B$ scale with
$N$ and $k$, respectively. The crucial identity
\begin{equation}\label{2.5.2}
 \tr K^m=\Str B^m,\quad\forall m\in\mathbb{N},
\end{equation}
yields the supersymmetric extension of the rotation invariant characteristic function,
\begin{equation}\label{2.5.3}
 \Phi(K)=\Phi(\tr K,\tr K^2,...)=\Phi(\Str B,\Str B^2,...)=\Phi(B) \ ,
\end{equation}
which is now viewed as a function in ordinary and superspace.
We rewrite it by inserting a proper Dirac--distribution in superspace,
\begin{eqnarray}
 \Phi(B) & = & \int \Phi(\rho)\delta(\rho-B)d[\rho]\\
 & \sim & \int\int \Phi(\rho)\exp[\imath\Str(\rho-B)\sigma]d[\rho]d[\sigma]\label{2.5.4} \ ,
\end{eqnarray}
where the supermatrix $\rho$ and $\sigma$ are introduced as
integration variables. The vectors of ordinary and Grassmann variables
now appear as in the conventional Hubbard--Stratonovich transformation
and can hence be integrated out in the same way. We are left with the
integrals over $\rho$ and $\sigma$. If we do the integral over $\rho$
we arrive at the result
\begin{equation}\label{2.5.5}
 Z_k\left(x^{-}+J\right)\sim\int Q(\sigma)\Sdet^{-N/\gamma_1}(\sigma-x^--J)d[\sigma].
\end{equation}
for the generating function.  The superfunction $Q$ is the
superspace Fourier transform of $\Phi$ and plays the role of a
probability density in superspace,
\begin{equation}\label{2.5.6}
 Q(\sigma)=\int \Phi(\rho)\exp(\imath\Str\rho\sigma)d[\rho] \ .
\end{equation}
If we choose to integrate over $\sigma$ instead, we obtain another representation
of the generating function
\begin{equation}\label{2.5.7}
 Z_k\left(x^{-}+J\right)\sim\int \Phi(\rho)I(\rho)\exp[-\imath\Str\rho(x^-+J)]d[\rho] \ ,
\end{equation}
which still contains the characteristic function. The distribution
$I(\rho)$ appears.  It is the supersymmetric version of the
Ingham--Siegel integral. It is a rotation invariant function resulting
from the Fourier transformation of the superdeterminant in
Eq.~\eref{2.5.5}. 

One way to proceed further is to diagonalize the supermatrix $\rho$
and to integrate over the angles. We may omit Efetov--Wegner terms and
have
\begin{equation}\label{2.5.8}
 Z_k\left(x^{-}+J\right)\sim\int \Phi(r)I(r)\varphi(-\imath r,x^-+J)d[r],
\end{equation}
where $\varphi$ is a supermatrix Bessel--function. The differentiation
with respect to $J$ gives $\widehat{R}_k$. We can introduce other
signatures of $L$ by Fourier transformation of Eq. \eref{2.5.7} and
identification with Eq. \eref{2.6}. Eventually, we find the
correlation functions $R_k$.

\section{Generalized Hubbard--Stratonovich transformation}\label{sec3}

In Sec. \ref{sec3.1}, we express the determinants in Eq. \eref{2.8} as Gaussian integrals and introduce the characteristic function of the matrix ensemble. In Sec. \ref{sec3.2}, we qualitatively present the duality between ordinary and superspace which is quantitatively discussed in Sec. \ref{sec3.3}. Then, we restrict the matrix ensembles to the classical symmetry classes. In Sec. \ref{sec3.4}, we investigate the diagonalization of the dyadic matrix $K$ appearing from the Gaussian integrals. The ambiguity of the supersymmetric extension of the characteristic function is discussed in Sec. \ref{sec3.5}. In Sec. \ref{sec3.6}, we present the symmetries of the appearing supermatrices. In Sec. \ref{sec3.7}, we replace the dyadic supermatrix in the supersymmetric extended characteristic function with a symmetric supermatrix discussed in the section before.

\subsection{Average over the ensemble and the characteristic function}\label{sec3.1}

To formulate the generating function as a supersymmetric integral, we consider a complex Grassmann algebra $\Lambda=\bigoplus\limits_{j=0}^{2Nk}\Lambda_j$ with $Nk$-pairs $\{\zeta_{jp},\zeta_{jp}^*\}_{j,p}$ of Grassmann variables \cite{Ber87}. We define the $k$ anticommuting vectors and their adjoint
\begin{equation}\label{3.1}
 \zeta_p=(\zeta_{1p},\ldots,\zeta_{Np})^T\ \ \ {\rm and}\ \ \ \zeta_p^\dagger=(\zeta_{1p}^*,\ldots,\zeta_{Np}^*)\ ,
\end{equation}
respectively. For integrations over Grassmann variables, we use the conventions of Ref.  \cite{KKG08}.
We also consider $k$\ $N$--dimensional complex vectors $\{z_p,z_p^{\dagger}\}_{1\leq p\leq k}$. In the usual way, we write the determinants as Gaussian integrals and find for Eq. \eref{2.8}
\begin{eqnarray}\fl
  Z_k(x^-+J) & = & (-\imath)^{Nk}\int\limits_{\mathfrak{M}_N}\int\limits_{\mathfrak{C}_{kN}}d[\zeta]d[z]d[H]P(H)\times\nonumber\\
  & \times &{\rm exp}\left(\imath\sum\limits_{p=1}^k\left\{\zeta_p^\dagger[H-(x_p^-+J_p)\eins_N]\zeta_p+ z_p^\dagger[H-(x_p^--J_p)\eins_N]z_p\right\}\right)\label{3.5}
\end{eqnarray}
where $d[\zeta]=\prod\limits_{p=1}^k\prod\limits_{j=1}^Nd\zeta_{jp}d\zeta_{jp}^*$, $d[z]=\prod\limits_{p=1}^k\prod\limits_{j=1}^Ndz_{jp}dz_{jp}^*$ and $\mathfrak{C}_{kN}=\mathbb{C}^{kN}\times\Lambda_{2Nk}$. Using
\begin{equation}\label{3.6}
 \sum\limits_{p=1}^k\left(\zeta_p^\dagger H\zeta_p+z_p^\dagger Hz_p\right)=\tr H\widetilde{K}
\end{equation}
with
\begin{equation}\label{3.7}
 \widetilde{K}=\sum\limits_{p=1}^k\left(z_pz_p^\dagger-\zeta_p\zeta_p^\dagger\right)
\end{equation}
leads to
\begin{eqnarray}\fl
  Z_k(x^-+J) & = & (-\imath)^{Nk}\int\limits_{\mathfrak{C}_{kN}}\mathcal{F}P\left(\hat{\pi}(\mathfrak{M}_N;\widetilde{K})\right)\times\nonumber\\
  & \times&{\rm exp}\left(-\imath\sum\limits_{p=1}^k\left[(x_p^-+J_p)\zeta_p^\dagger\zeta_p +(x_p^--J_p)z_p^\dagger z_p\right]\right)d[\zeta]d[z]\ .\label{3.8}
\end{eqnarray}
where the integration over $H$ is the Fourier transformation of the probability density $P$,
\begin{equation}\label{3.9}
 \mathcal{F}P\left(\hat{\pi}(\mathfrak{M}_N;\widetilde{K})\right)=\int\limits_{\mathfrak{M}_N}P(H)\exp\left(\imath\tr H\widetilde{K}\right)d[H]\ .
\end{equation}
This Fourier transform is called characteristic function and is denoted by $\Phi$ in Ref.  \cite{Guh06} and in Eq. \eref{2.5.1}. The projection operator $\hat{\pi}(\mathfrak{M}_N)$ onto the space $\mathfrak{M}_N$ is crucial. For $\mathfrak{M}_{\gamma_2N}=\Herm(\beta,N)$ the projection operator is
\begin{equation}\label{3.10}
 \hat{\pi}\left(\Herm(\beta,N);\widetilde{K}\right)=\frac{1}{2}\left[\widetilde{K}+\widehat{Y}(\widetilde{K})\right]
\end{equation}
with
\begin{equation}\label{3.11}
 \widehat{Y}(\widetilde{K})=\left\{\begin{array}{ll}
            \widetilde{K}^T & ,\ \beta=1\\
	    \widetilde{K} & ,\ \beta=2\\
	     \left(Y_{{\rm s}}\otimes\eins_N\right)\widetilde{K}^T\left(Y_{{\rm s}}^T\otimes\eins_N\right) & ,\ \beta=4
           \end{array}\right.
\end{equation}
and the symplectic unit
\begin{equation}\label{3.11b}
 Y_{{\rm s}}=\left[\begin{array}{cc} 0 & 1 \\ -1 & 0 \end{array}\right]\ ,
\end{equation}
where $\eins_N$ is the $N\times N$--unit matrix. The transposition in Eq. \eref{3.11} can also be replaced by the complex conjugation due to $\widetilde{K}^\dagger=\widetilde{K}$. The projection onto the set of diagonal matrices $\bigoplus\limits_{j=1}^N\mathbb{R}$ is
\begin{equation}\label{3.12}
 \hat{\pi}\left(\bigoplus_{j=1}^N\mathbb{R};\widetilde{K}\right)=\diag\left(\widetilde{K}_{11},\widetilde{K}_{22},\ldots,\widetilde{K}_{NN}\right) \ .
\end{equation}

\subsection{Duality between ordinary and superspace}\label{sec3.2}

Is it always possible to find a supermatrix representation for the characteristic function $\mathcal{F}P$ such that Eq.~\eref{3.8} has an integral representation over supermatrices as it is known \cite{Guh06,LSZ07} for rotation invariant $P$ on $\mathfrak{M}_{\gamma_2N}=\Herm(\beta,N)$? The integral \eref{3.8} is an integral over the supervectors $v_j=(z_{j1}^*,\ldots,z_{jk}^*,-\zeta_{j1}^*,\ldots,-\zeta_{jk}^*)^T$ and their adjoint $v_j^\dagger=(z_{j1},\ldots,z_{jk},\zeta_{j1},\ldots,\zeta_{jk})$. The adjoint ``$\dagger$'' is the complex conjugation with the supersymmetric transposition and ``$T$'' is the ordinary transposition. The entries of the matrix $\widetilde{K}$ are $v_n^\dagger v_m$. If we do not use any symmetry of the matrix ensemble, we can write these scalar products of supervectors as supertraces
\begin{equation}\label{3.13}
 v_n^\dagger v_m=\Str v_mv_n^\dagger\ .
\end{equation}
Then, we can transform each of these supertraces with a Dirac--distribution to an integral over a $(k+k)\times(k+k)$--supermatrix. We defined the Dirac--distribution in superspace as in Refs.  \cite{DeW84,BasAke07}. The ambiguity discussed in Ref.  \cite{BEKYZ07} occurring by such a transformation is discussed in the subsections \ref{sec3.5} and \ref{sec5.3}.

The procedure above is tedious. Using the symmetries of the ensemble ($\mathcal{F}P,\mathfrak{M}_N$), we can reduce the number of integrals in superspace. We will see that the number of commuting real integrals and of Grassmannian integrals is  $2k^2+2k^2$ ($\beta=2$) or $4k^2+4k^2$ ($\beta\in\{1,4\}$) for a rotation invariant matrix ensembles on $\Herm(\beta,N)$. If there is not a symmetry the number of integrals has not been reduced. One has to integrate over $N(N+1)$ ordinary hermitian $k\times k$--matrices and their corresponding anticommuting parameters if the transformation above is used.

\subsection{Analysis of the duality between ordinary and superspace}\label{sec3.3}

We consider an orthonormal basis $\{A_n\}_{1\leq n\leq d}$ of $\mathfrak{M}_N$ where $d$ is the dimension of $\mathfrak{M}_N$. We use the trace $\tr{A_nA_m}=\delta_{nm}$ as the scalar product and recall that $\mathfrak{M}_N$ is a real vector space. Every element of this basis is represented as
\begin{equation}\label{3.14}
 A_n=\sum\limits_{j=1}^N\lambda_{jn}e_{jn}e_{jn}^\dagger\ \ \ {\rm with}\ \ \ \sum\limits_{j=1}^N\lambda_{jn}^2=1\ .
\end{equation}
Here, $e_{jn}$ are the normalized eigenvectors of $A_n$ to the eigenvalues $\lambda_{jn}$. Then, we construct every matrix $H\in \mathfrak{M}_N$ in this basis
\begin{equation}\label{3.15}
 H=\sum\limits_{n=1}^d h_nA_n\ .
\end{equation}
We find for the characteristic function
\begin{eqnarray}\fl
  \mathcal{F}P\left(\hat{\pi}(\mathfrak{M}_N;\widetilde{K})\right) & = & \int\limits_{\mathfrak{M}_N}P\left(\sum\limits_{n=1}^d h_nA_n\right){\rm exp}\left(\imath\sum\limits_{n=1}^d h_n\tr A_n\widetilde{K}\right)d[H]=\nonumber\\
  & = & \mathcal{F}P\left(\sum\limits_{n=1}^d \tr\left(\widetilde{K}A_n\right)A_n\right)\ .\label{3.16}
\end{eqnarray}
With help of Eq. \eref{3.14} and an equation analogous to \eref{3.13}, the characteristic function is
\begin{equation}\label{3.17}
 \mathcal{F}P\left(\hat{\pi}(\mathfrak{M}_N;\widetilde{K})\right)=\mathcal{F}P\left(\sum\limits_{n=1}^d\Str\left(\sum\limits_{j=1}^N\lambda_{jn}Ve_{jn}e_{jn}^\dagger V^\dagger\right) A_n\right)
\end{equation}
with $V=(v_1,\ldots, v_N)$. We see that the matrix $\widetilde{K}$ is projected onto
\begin{equation}\label{3.17b}
 K=\hat{\pi}(\mathfrak{M}_N;\widetilde{K})
\end{equation}
where the projection is the argument of the characteristic function in Eq. \eref{3.16}. The matrices in the supertraces of \eref{3.17} can be exchanged by $(k+k)\times(k+k)$--supermatrices with the Delta--distributions described above. If the ensemble has no symmetry then we have reduced the number of supermatrices to the dimension of $\mathfrak{M}_N$. Nevertheless, we can find a more compact supersymmetric expression of the matrix $K$ such that the number of the resulting integrals only depends on $k$ but not on $N$. This is possible if $K$ is a dyadic matrix of vectors where the number of vectors is independent of $N$ and the probability distribution only depends on invariants of $H$. The ensembles with $\mathfrak{M}_{\gamma_2N}=\Herm(\beta,N)$ and a probability density $P$ invariant under the action of $\U^{(\beta)}(N)$ fulfil these properties. It is known \cite{Guh06,LSZ07} that these cases have a very compact supersymmetric expression. Furthermore, these ensembles are well analyzed for Gaussian--distributions with help of the Hubbard--Stratonovitch transformation \cite{Efe83,Efe97,VWZ85}.

In the present context, the cases of interest are $\mathfrak{M}_{\gamma_2N}=\Herm(\beta,N)$ with a probability density $P$ invariant under the action $\U^{(\beta)}(N)$. We need this symmetry to simplify Eq. \eref{3.17}. Let $N\geq\gamma_1k$. This restriction also appears in the superbosonization formula \cite{LSZ07}. If $N<\gamma_1k$, one has to be modify the calculations below. For the superbosonization formula, Bunder, Efetov, Kravtsov, Yevtushenko, and Zirnbauer \cite{BEKYZ07} presented such a modification.

The symmetries of a function $f$ carry over to its Fourier transform $\mathcal{F}f$. Thus, the characteristic function $\mathcal{F}P$ is invariant under the action of $\U^{(\beta)}(N)$. Let $\widetilde{K}_0$ be an arbitrary ordinary hermitian matrix in the Fourier transformation \eref{3.9} of the probability density. We assume that the characteristic function is analytic in the eigenvalues of $\widetilde{K}_0$. Then, we expand $\mathcal{F}P$ as a power series in these eigenvalues. Since the characteristic function is rotation invariant every single polynomial in this power series of a homogeneous degree is permutation invariant. With help of the fundamental theorem of symmetric functions \cite{Wae71} we rewrite these polynomials in the basis of elementary polynomials. This is equivalent to writing these polynomials in the basis of the traces $\tr\left[\hat{\pi}\left(\Herm(\beta,N),\widetilde{K}_0\right)\right]^m$, $m\in\mathbb{N}$. The analytic continuation of $\mathcal{F}P$ from $\widetilde{K}_0$ to $\widetilde{K}$ yields that the characteristic function in \eref{3.9} only depends on $\tr\left[\hat{\pi}\left(\Herm(\beta,N),\widetilde{K}\right)\right]^m$, $m\in\mathbb{N}$.

Defining the matrix
\begin{equation}\label{3.18}
 V^\dagger=(z_1,\ldots,z_k,Yz_1^*,\ldots,Yz_k^*,\zeta_1,\ldots,\zeta_k,Y\zeta_1^*,\ldots,Y\zeta_k^*)
\end{equation}
and its adjoint
\begin{equation}\label{3.19}
 V=(z_1^*,\ldots,z_k^*,Yz_1,\ldots,Yz_k,-\zeta_1^*,\ldots,-\zeta_k^*,Y\zeta_1,\ldots,Y\zeta_k)^T
\end{equation}
with
\begin{equation}\label{3.20}
 Y=\left\{\begin{array}{ll}
            \eins_N & ,\ \beta=1\\
	    0 & ,\ \beta=2\\
	     Y_{{\rm s}}^T\otimes\eins_N & ,\ \beta=4
           \end{array}\right.,
\end{equation}
we find
\begin{equation}\label{3.21}
 K=\hat{\pi}\left(\Herm(\beta,N);\widetilde{K}\right)=\frac{1}{\tilde{\gamma}}V^\dagger V\ .
\end{equation}
The crucial identity
\begin{equation}\label{3.22}
 \tr(V^\dagger V)^m=\Str(VV^\dagger)^m
\end{equation}
holds for all $\beta$. It connects ordinary and superspace. For $\beta=2$, a proof can be found in Ref.  \cite{Guh06}. In \ref{app1}, we show that the equation
\begin{equation}\label{3.22b}
 \Str V_1V_2=\Str V_2V_1
\end{equation}
holds for all rectangular matrices of the form
\begin{equation}\label{3.23}
 V_1=\left[\begin{array}{cc} \overbrace{A_1}^{a} & \overbrace{B_1}^{b}\hspace*{0.3mm}\}c \\ C_1 & D_1\hspace{2.1mm}\}d\end{array}\right]\ \ \ {\rm and}\ \ \ V_2=\left[\begin{array}{cc} \overbrace{A_2}^{c} & \overbrace{B_2}^{d}\hspace*{0.3mm}\}a \\ C_2 & D_2\hspace*{2mm}\}b\end{array}\right]
\end{equation}
where $A_j$ and $D_j$ have commuting entries and $B_j$ and $C_j$ anticommuting ones. This implies in particular that Eq. \eref{3.22} holds for all $\beta$. Hence, we reduced the amount of supermatrices corresponding to $\widetilde{K}$ in Eq. \eref{3.17} to one $(2k+2k)\times(2k+2k)$--supermatrix. In Ref.  \cite{Guh06}, the characteristic function $\Phi$ was, with help of Eq. \eref{3.22}, extended to superspace. We follow this idea and, then, proceed with the Dirac--distribution mentioned above.

\subsection{Problems when diagonalizing $K$}\label{sec3.4}

In Ref.  \cite{Guh06}, two approaches of the duality relation between ordinary and superspace were presented. The first approach is the duality equation \eref{3.22} for $\beta=2$. In our article, we follow this idea. In the second approach, the matrix $K$ was diagonalized. With the eigenvalues of $K$, a projection operator was constructed for the definition of a reduced probability density according to the probability density $P$.

The latter approach fails because $K$ is only diagonalizable if it has no degeneracy larger than $\gamma_2$. Moreover for diagonalizable  $K$, one can not find an eigenvalue $\lambda=0$. This is included in the following statement which we derive in \ref{app5}.
\begin{theorem}\label{t0}\ \\
 Let $N,\widetilde{N}\in\mathbb{N}$, $H^{(0)}\in\Herm(\beta,N)$, $l\in\mathbb{R}^{\widetilde{N}}$ and $\{\tau_q\}_{1\leq q\leq \widetilde{N}}$ $\gamma_2N$--dimensional vectors consisting of Grassmann variables $\tau_q=(\tau_q^{(1)},\ldots,\tau_q^{(\gamma_2N)})^T$. Then, the matrix
\begin{equation}\label{t0.1}
 H=H^{(0)}+\sum\limits_{q=1}^{\widetilde{N}}l_q\left[\tau_q\tau_q^\dagger+\widehat{Y}\left(\tau_q^*\tau_q^T\right)\right]
\end{equation}
can not be diagonalized $H=U\diag(\lambda_1,\ldots,\lambda_N)U^\dagger$ by a matrix $U$ with the properties
\begin{equation}\label{t0.2}
 U^\dagger U=UU^\dagger=\eins_N\ ,\ \ U^*=\widehat{Y}(U)
\end{equation}
and the body of $U$ lies in $\U^{(\beta)}(N)$ iff $H^{(0)}$ has degeneracy larger than $\gamma_2$. Moreover, $H$ has no eigenvalue $\lambda\in\mathbb{R}$.
\end{theorem}

In our particular case, $K$ can not be diagonalized for $k<N-1$. Hence, we do not follow the second approach of Ref.  \cite{Guh06}. We emphasize that none of the other results in Ref.  \cite{Guh06} is affected as they are proven by the correct first approach which we pursue here.

\subsection{Ambiguity of the characteristic function in the supersymmetric extension}\label{sec3.5}

In this section, we discuss the problem that the extension of the characteristic function $\mathcal{F}P$ from ordinary matrices to supermatrices is not unique. This results from the fact that symmetric supermatrices comprise two kinds of eigenvalues, i.e. bosonic and fermionic eigenvalues. Whereas ordinary symmetric matrices have only one kind of eigenvalues. In the supertraces, these two different kinds are differently weighted by a minus sign. To illustrate this problem, we also give a simple example.

The rotation invariance of $\mathcal{F}P$ enables us to choose a representation $\mathcal{F}P_0$ of $\mathcal{F}P$ acting on an arbitrary number of matrix invariants
\begin{equation}\label{3.25}
 \mathcal{F}P_0\left(\tr K^m|m\in\mathbb{N}\right)=\mathcal{F}P(K)\ .
\end{equation}
For this representation, a unique superfunction exists defined by
\begin{equation}\label{3.26}
 \Phi_0(\sigma)=\mathcal{F}P_0\left(\Str\sigma^m|m\in\mathbb{N}\right)
\end{equation}
where
\begin{equation}\label{3.h1}
 \mathcal{F}P_0\left(\Str B^m|m\in\mathbb{N}\right)=\mathcal{F}P_0\left(\tr K^m|m\in\mathbb{N}\right)
\end{equation}
with $B=\tilde{\gamma}^{-1}VV^\dagger$. However, the choice of the representation $\mathcal{F}P_0$ is not unique. The question arises whether it is a well defined object. It is clear that two representations $\mathcal{F}P_0$ and $\mathcal{F}P_1$ are equal on $\Herm(\beta,N)$ due to the Cayley--Hamilton theorem,
\begin{equation}\label{3.27a}
 \mathcal{F}P_0(H)=\mathcal{F}P_1(H)\ ,\ H\in\Herm(\beta,N).
\end{equation}
The Cayley--Hamilton theorem states that there is a polynomial which is zero for $H$. Thus, $H^M$ with $M>N$ is a polynomial in $\{H^{n}\}_{1\leq n\leq N}$. Plugging an arbitrary symmetric supermatrix $\sigma$ into the corresponding superfunctions $\Phi_0$ and $\Phi_1$ we realize that the choices are not independent such that
\begin{equation}\label{3.27}
 \Phi_0(\sigma)\neq\Phi_1(\sigma)
\end{equation}
holds for some $\sigma$.

For example with $N=2$, $k=1$ and $\beta=2$, let the characteristic function $\mathcal{F}P(H)=\mathcal{F}P_0\left(\tr H^3\right)$. We get with help of the Cayley--Hamilton theorem
\begin{equation}\fl\label{3.28}
 \mathcal{F}P_1\left(\tr H^2,\tr H\right)=\mathcal{F}P_0\left(2\tr H\tr H^2-\tr^3 H\right)=\mathcal{F}P_0\left(\tr H^3\right)=\mathcal{F}P(H)\ .
\end{equation}

Let the set of $\U^{(\beta)}(p/q)$--symmetric supermatrices be
\begin{eqnarray}
 \left\{\sigma\in{\rm Mat}(\tilde{\gamma}p/\tilde{\gamma}q)\left|\sigma^\dagger=\sigma,\ \sigma^*=\widehat{Y}_{{\rm S}}(\sigma)
\right.\right\}{\rm\ and}\label{3.28b}\\
 \widehat{Y}_{{\rm S}}(\sigma)=\left\{\begin{array}{ll}
   \left[\begin{array}{cc} \eins_{2p} & 0 \\ 0 & Y_{{\rm s}}\otimes\eins_q \end{array}\right]\sigma\left[\begin{array}{cc} \eins_{2p} & 0 \\ 0 & Y_{{\rm s}}^T\otimes\eins_q \end{array}\right] & ,\ \beta=1, \\
   \sigma^* & ,\ \beta=2, \\
   \left[\begin{array}{cc} Y_{{\rm s}}\otimes\eins_p & 0 \\ 0 & \eins_{2q} \end{array}\right]\sigma\left[\begin{array}{cc} Y_{{\rm s}}^T\otimes\eins_p & 0 \\ 0 & \eins_{2q} \end{array}\right] & ,\ \beta=4,
  \end{array}\right.\label{3.28c}
\end{eqnarray}
with respect to the supergroups
\begin{equation}\label{5.1}
 \U^{(\beta)}(p/q)=\left\{\begin{array}{ll}
                 \UOSp^{(+)}(p/2q) & ,\ \beta=1\\
                 \U(p/q) & ,\ \beta=2\\
                 \UOSp^{(-)}(2p/q) & ,\ \beta=4
                \end{array}\right.\ .
\end{equation}
${\rm Mat}(\tilde{\gamma}p/\tilde{\gamma}q)$ is the set of $(\tilde{\gamma}p+\tilde{\gamma}q)\times(\tilde{\gamma}p+\tilde{\gamma}q)$--supermatrices with the complex Grassmann algebra $\bigoplus\limits_{j=0}^{8k^2}\Lambda_j$. The definition of the two representations $\UOSp^{(\pm)}$ of the supergroup $\UOSp$ can be found in Refs.  \cite{KohGuh05,KKG08}. We refer to the classification of Riemannian symmetric superspaces by Zirnbauer \cite{Zir96}.

We consider a $\U(1/1)$--symmetric supermatrix $\sigma$. This yields for the supersymmetric extension of Eq. \eref{3.28}
\begin{equation}\fl\label{3.29}
 \mathcal{F}P_0\left(2\Str\sigma\Str\sigma^2-\Str^3\sigma\right)\neq\mathcal{F}P_0\left(\Str\sigma^3\right)=\mathcal{F}P_0\left(\frac{1}{4}\left(3\frac{\Str^2\sigma^2}{\Str\sigma}+\Str^3\sigma\right)\right)\ .
\end{equation}
One obtains the last equation with a theorem similar to the Cayley--Hamilton theorem. More specificly, there exists a unique polynomial  equation of order two
\begin{equation}\label{3.30}
 \sigma^2-\frac{\Str\sigma^2}{\Str\sigma}\sigma-\frac{1}{4}\left(\Str^2\sigma-\frac{\Str^2\sigma^2}{\Str^2\sigma}\right)=0\ ,
\end{equation}
for a $\U(1/1)$--symmetric supermatrix $\sigma$.

The resulting integral in Sec. \ref{sec4} for the generating function $Z_k|_{\mathfrak{M}_N=\Herm(\beta,N)}$ is invariant under the choice of $\Phi_0$. This is proven in Sec. \ref{sec5.3}. Such an ambiguity of the supersymmetric extension of the characteristic function was also investigated by the authors of Ref.  \cite{BEKYZ07}. They avoided the question of the definition of a Dirac--distribution on superspace by the superbosonization formula. We introduce for the supersymmetric extension from Eq. \eref{3.h1} to Eq. \eref{3.26} a Dirac--distribution depending on the representation of the superfunction.

\subsection{Symmetries of the supermatrices}\label{sec3.6}

We find for a chosen representation $\mathcal{F}P_0$
\begin{equation}\label{3.31}
 Z_k(x^-+J)=(-\imath)^{k_2N}\int\limits_{\mathfrak{C}^{k_2N}}\Phi_0(B) \exp\left[-\imath\Str (x^-+J)B\right]d[\zeta]d[z]\ .
\end{equation}
Here, we introduce $k_2=\gamma_2k$, $k_1=\gamma_1k$ and $\tilde{k}=\tilde{\gamma}k$. We will simplify the integral \eref{3.31} to integrals over $k_1$ eigenvalues in the Boson--Boson block and over $k_2$ eigenvalues in the Fermion--Fermion block.

For every $\beta$, we have
\begin{equation}\label{3.32}
 B^\dagger=B\ ,
\end{equation}
i.e. $B$ is self-adjoint. The complex conjugation yields
\begin{equation}\label{3.33}
 B^*=\left\{\begin{array}{ll}
                \widetilde{Y}B\widetilde{Y}^T\qquad,\ \beta\in\{1,4\} \\
		\widetilde{Y}B^*\widetilde{Y}^T\qquad,\ \beta=2
               \end{array}\right.
\end{equation}
with the $(2k+2k)\times(2k+2k)$--supermatrices
\begin{equation}\fl\label{3.34}
 \left.\widetilde{Y}\right|_{\beta=1}=\left[\begin{array}{ccc} 0 & \eins_k & 0  \\ \eins_k & 0 & 0  \\ 0 & 0 & Y_{{\rm s}}\otimes\eins_k \end{array}\right]\qquad ,\qquad \left.\widetilde{Y}\right|_{\beta=4}=\left[\begin{array}{ccc} Y_{{\rm s}}\otimes\eins_k & 0 & 0 \\ 0 & 0 & \eins_k \\ 0 & \eins_k & 0 \end{array}\right]
\end{equation}
and $\left.\widetilde{Y}\right|_{\beta=2}=\diag(1,0,1,0)\otimes\eins_k$. We notice that for the unitary case $B$ is effectively a $(k+k)\times(k+k)$--supermatrix, i.e. half the dimension. With help of the properties \eref{3.32} and \eref{3.33} we construct the supermatrix sets
\begin{equation}\fl\label{3.35}
 \widetilde{\Sigma}_{0}(\beta,k)=\left\{\sigma\in{\rm Mat}(2k/2k)\left|\sigma^\dagger=\sigma,\ \sigma^*=\left\{\begin{array}{ll}
                \widetilde{Y}\sigma\widetilde{Y}^T &,\ \beta\in\{1,4\} \\
		\widetilde{Y}\sigma^*\widetilde{Y}^T &,\ \beta=2
               \end{array}\right\}\right.\right\}\ .
\end{equation}
A matrix in $\widetilde{\Sigma}_{0}(\beta,k)$ fulfils the odd symmetry \eref{3.33}. We transform this symmetry with the unitary transformations
\begin{equation}\fl\label{3.36}
 U|_{\beta=1}=\frac{1}{\sqrt{2}}\left[\begin{array}{ccc} \eins_{k} & \eins_{k} & 0 \\ -\imath\eins_{k} & \imath\eins_{k} & 0 \\ 0 & 0 & \sqrt{2}\ \eins_{2k} \end{array}\right]\ \ ,\ \ U|_{\beta=4}=\frac{1}{\sqrt{2}}\left[\begin{array}{ccc} \sqrt{2}\ \eins_{2k} & 0 & 0 \\ 0 & \eins_{k} & \eins_{k}\\ 0 & -\imath\eins_{k} & \imath\eins_{k} \end{array}\right],
\end{equation}
$U|_{\beta=2}=\eins_{4k}$, according to the Dyson--index, arriving at the well--known symmetries of symmetric supermatrices \cite{Zir96}, see also Eq. \eref{3.28b}. Defining the sets $\Sigma_{0}(\beta,k)=U\widetilde{\Sigma}_{0}(\beta,k)U^\dagger$, we remark that the body of the Boson--Boson block of any element in these sets is a matrix in $\Herm(\beta,k_1)$. The body of the Fermion--Fermion block of any matrix in $\Sigma_{0}(\beta,k)$ lies in $\Herm(4/\beta,k_2)$.

We introduce a generalized Wick--rotation $e^{\imath\psi}$ to guarantee the convergence of the supermatrix integrals. The usual choice of a Wick--rotation is $e^{\imath\psi}=\imath$ for investigations of Gaussian probability densities \cite{Guh06,Efe83,VWZ85}. Here, general Wick--rotations \cite{KKG08} are also of interest. Probability densities which lead to superfunction as $\exp\left(-\Str\sigma^4\right)$ do not converge with the choice $\imath$. Thus, we consider the modified sets
\begin{equation}\label{3.37}
 \Sigma_{\psi}(\beta,k)=\widehat{\Psi}_\psi\Sigma_{0}(\beta,k)\widehat{\Psi}_\psi\ .
\end{equation}
with $\widehat{\Psi}_\psi=\diag(\eins_{2k},e^{\imath\psi/2}\eins_{2k})$. Let $\Sigma_{\psi}^0(\beta,k)$ be the set of supermatrices which contains only zero and first order terms in the Grassmann variables.

In the sequel, we restrict our calculations to superfunctions which possess a Wick--rotation such that the integrals below are convergent. We have not further explored the set of superfunctions with this property, but we know that this set has to be very large and sufficient for our purposes. For example, superfunctions of the form
\begin{equation}\label{3.37b}
 \Phi_0(\sigma)=\widetilde{\Phi}(\sigma)\exp\left(-\Str\sigma^{2n}\right),\quad n\in\mathbb{N},
\end{equation}
fulfil this property if ${\rm ln}\widetilde{\Phi}(\sigma)$ does not increase as fast as $\Str\sigma^{2n}$ at infinity.

\subsection{Transformation to supermatrices by a Dirac--distribution}\label{sec3.7}

Following Refs.  \cite{LSSS95,Guh06,BasAke07}, $\Phi_0(B)$ can be written as a convolution in the space of supermatrices $\Sigma_{\psi}^0(\beta,k)$ with a Dirac--distribution. We have
\begin{eqnarray}
 Z_k(x^-+J)& = & (-\imath)^{k_2N}\int\limits_{\mathfrak{C}_{k_2N}}\int\limits_{\Sigma_{\psi}^0(\beta,k)}\Phi_0(\rho)\delta\left(\rho-UBU^\dagger\right)d[\rho]\times\nonumber\\
 & \times & \exp\left[-\imath\Str (x^-+J)B\right]d[\zeta]d[z]\label{3.38}
\end{eqnarray}
where the measure is defined as
\begin{equation}\label{3.39}
 d[\rho]=d[\rho_1]d[\rho_2]\underset{1\leq n\leq k_1}{\prod\limits_{1\leq m\leq k_2} }d\eta_{nm}d\eta_{nm}^*\ .
\end{equation}
Here, $\{\eta_{nm},\eta_{nm}^*\}$ are pairs of generators of a Grassmann algebra, while $\rho_1$ is the Boson--Boson and $\rho_2$ is the Fermion--Fermion block without the phase of the Wick--rotation. Since $\rho_1$ and $\rho_2$ are in $\Herm(\beta,k_1)$ and $\Herm(4/\beta,k_2)$, respectively, we use the real measures for $d[\rho_1]$ and $d[\rho_2]$ which are defined in Ref.  \cite{KKG08}. We exchange the Dirac--distribution by two Fourier transformations as in Refs.  \cite{Guh06,BasAke07}. Then, Eq. \eref{3.38} becomes
\begin{eqnarray}
  Z_k(x^-+J)& = & (-\imath)^{k_2N}2^{2k(k-\tilde{\gamma})}\int\limits_{\mathfrak{C}_{k_2N}}\int\limits_{\Sigma_{-\psi}^0(\beta,k)}\mathcal{F}\Phi_0(\sigma)\times\nonumber\\
  & \times & \exp\left[\imath\Str B\left(U^\dagger\sigma U-x^--J\right)\right]d[\sigma]d[\zeta]d[z]\label{3.40}
\end{eqnarray}
where the Fourier transform of $\Phi_0$ is
\begin{equation}\label{3.41}
 \mathcal{F}\Phi_0(\sigma)=\int\limits_{\Sigma_{\psi}^0(\beta,k)}\Phi_0(\rho)\exp\left(-\imath\Str\rho\sigma\right)d[\rho]\ .
\end{equation}
We write the supertrace in the exponent in Eq. \eref{3.40} as a sum over expectation values
\begin{equation}\label{3.42}
 \Str B\left(U^\dagger\sigma U-x^- -J\right)=\frac{1}{\tilde{\gamma}}\sum\limits_{j=1}^{N}\tr\Psi_j^\dagger\left(U^\dagger\sigma U-x^- -J\right)\Psi_j
\end{equation}
with respect to the real, complex or quaternionic supervectors
\begin{equation}\fl\label{3.43}
 \Psi_j^\dagger=\left\{\begin{array}{ll}
                        \left\{z_{jn},z_{jn}^*,\zeta_{jn},\zeta^*_{jn}\right\}_{1\leq n\leq k} & \hspace*{-0.1cm},\ \beta=1\\
                        \left\{z_{jn},0,\zeta_{jn},0\right\}_{1\leq n\leq k} & \hspace*{-0.1cm},\ \beta=2\\
                        \left\{\left[\begin{array}{c} z_{jn} \\ z_{j+N,n} \end{array}\right],\left[\begin{array}{c} -z_{j+N,n}^* \\ z_{jn}^* \end{array}\right],\left[\begin{array}{c} \zeta_{jn} \\ \zeta_{j+N,n} \end{array}\right],\left[\begin{array}{c} -\zeta_{j+N,n}^* \\ \zeta_{jn}^* \end{array}\right]\right\}_{1\leq n\leq k} & \hspace*{-0.1cm},\ \beta=4
                       \end{array}\right.
\end{equation}
The integration over one of these supervectors yields
\begin{equation}\fl\label{3.44}
 \int\limits_{\mathfrak{C}_{k_2}}{\rm exp}\left[\frac{\imath}{\tilde{\gamma}}\tr\Psi_j^\dagger\left(U^\dagger\sigma U-x^- -J\right)\Psi_j\right]d[\Psi_j]= \imath^{k_2}\Sdet^{-1/\gamma_1}\mathfrak{p}\left(\sigma-x^- -J\right)\ .
\end{equation}
$\mathfrak{p}$ projects onto the non-zero matrix blocks of $\Sigma_{-\psi}(\beta,k)$ which are only $(k+k)\times (k+k)$--supermatrices for $\beta=2$. $\mathfrak{p}$ is the identity for $\beta\in\{1,4\}$. The Eq. \eref{3.44} is true because $U$ commutes with $x^-+J$. Then, Eq. \eref{3.40} reads
\begin{equation}\fl\label{3.45}
 Z_k(x^-+J)=2^{2k(k-\tilde{\gamma})} \int\limits_{\Sigma_{-\psi}^0(\beta,k)}\mathcal{F}\Phi_0(\sigma)\Sdet^{-N/\gamma_1}\mathfrak{p}\left(\sigma-x^- -J\right)d[\sigma]\ .
\end{equation}
Indeed, this result coincides with Ref.  \cite{Guh06} for $\beta=2$ where the Fourier transform $\mathcal{F}\Phi_0(\sigma)$ was denoted by $Q(\sigma)$. Eq. \eref{3.45} reduces for Gaussian ensembles with arbitrary $\beta$ to expressions as in Refs.  \cite{Efe97} and  \cite{VWZ85}. The integral is well defined because $\varepsilon$ is greater than zero and the body of the eigenvalues of the Boson--Boson block is real. The representation \eref{3.45} for the generating function can also be considered as a random matrix ensemble lying in the superspace.

Eq. \eref{3.45} is one reason why we called this integral transformation from the space over ordinary matrices to supermatrices as generalized Hubbard--Stratonovich transformation. If the probability density $P$ is Gaussian then we can choose $\Phi_0$ also as a Gaussian. Thus, this transformation above reduces to the ordinary Hubbard--Stratonovich transformation and the well-known result \eref{3.45}.

\section{The supersymmetric Ingham--Siegel integral}\label{sec4}

We perform a Fourier transformation in superspace for the convolution integral \eref{3.45} and find
\begin{equation}\label{4.1}\fl
 Z_k(x^-+J)=2^{2k(k-\tilde{\gamma})} \int\limits_{\Sigma_{\psi}^0(\beta,k)}\Phi_0(\rho)I_k^{(\beta,N)}(\rho)\exp\left[-\imath\Str\rho\left(x^{-}+J\right)\right]d[\rho]\ .
\end{equation}
Here, we have to calculate the supersymmetric Ingham--Siegel integral
\begin{equation}\label{4.2}
 I_k^{(\beta,N)}(\rho)=\int\limits_{\Sigma_{-\psi}^0(\beta,k)}\exp\left(-\imath\Str\rho\sigma^+\right)\Sdet^{-N/\gamma_1}\mathfrak{p}\sigma^+d[\sigma]
\end{equation}
with $\sigma^+=\sigma+\imath\varepsilon\eins_{4k}$.

Ingham \cite{Ing33} and Siegel \cite{Sie35} independently calculated a version of \eref{4.2} for ordinary real symmetric matrices. The case of hermitian matrices was discussed in Ref.  \cite{Fyo02}. Since we were unable to find the ordinary Ingham--Siegel integral also for the quaternionic case, we give the result here. It is related to Selbergs integral \cite{Meh91}. Let $R\in\Herm(\beta,m)$, $\varepsilon>0$ and a real number $n\geq m-1+2/\beta$, then we have
\begin{equation}\label{4.3}\fl
  \displaystyle\int\limits_{\Herm(\beta,m)}\exp\left(-\imath\tr RS^+\right){\det}^{-n/\gamma_1}S^+d[S]= \imath^{-\beta mn/2}G_{n-m,m}^{(\beta)} \displaystyle{\det}^{\lambda} R\ \Theta(R)
\end{equation}
where $S^+=S+\imath\varepsilon\eins_{\gamma_2m}$, the exponent is
\begin{equation}\label{c0}
 \lambda=\frac{n-m}{\gamma_1}-\frac{\gamma_1-\gamma_2}{2}
\end{equation}
and the constant is
\begin{equation}\label{c1}
 G_{n-m,m}^{(\beta)}=\left(\frac{\gamma_2}{\pi}\right)^{\beta m(n-m+1)/2-m}\prod\limits_{j=n-m+1}^{n}\frac{2\pi^{\beta j/2}}{\Gamma\left(\beta j/2\right)}\ .
\end{equation}
$\Gamma(.)$ is the Euler gamma--function and $\Theta(.)$ is the Heavyside-function for matrices which is defined as
\begin{equation}\label{4.4}
 \Theta(R)=\left\{\begin{array}{ll}
            	1	&	,\ R{\rm \ is\ positive\ definite}\\
		0	&	,\ {\rm else}
           \end{array}\right.\ .
\end{equation}
The ordinary Ingham--Siegel integral was recently used in the context of supersymmetry by Fyodorov \cite{Fyo02}. The integral was extended to the superspace $\Sigma_{\pi/2}^0(2,k)$ in Ref.  \cite{Guh06}. In this article, we need a generalization to all $\Sigma_{-\psi}^0(\beta,k)$, in particular $\beta=1,4$.

The integral \eref{4.2} is invariant under the action of $\U^{(\beta)}(k_1/k_2)$. Thus, it is convenient to consider $I(r,\varepsilon)$, where $r=\diag(r_{11},\ldots,r_{\tilde{k}1},r_{12},\ldots,r_{\tilde{k}2})$ is the diagonal matrix of eigenvalues of $\rho$ and contains nilpotent terms. The authors of Ref. \cite{BasAke07} claimed in their proof of Theorem 1 in Chapter 6 that the diagonalization at this point of the calculation yields Efetov--Wegner terms. These terms do not appear in the $\rho_2$ integration because we do not change the integration variables, i.e. the integration measure $d[\rho]$ remains the same. For the unitary case, see Ref.  \cite{Guh06}. We consider the eigenvalues of $\rho$ as functions of the Cartesian variables. We may certainly differentiate a function with respect to the eigenvalues if we keep track of how these differential operators are defined in the Cartesian representation.

As worked out in \ref{app3.1}, the supersymmetric Ingham--Siegel integral \eref{4.2} reads
\begin{equation}\fl\label{4.15}
  I_k^{(\beta,N)}(\rho)=\displaystyle C{\det}^\kappa r_1\Theta(r_1) {\det}^k r_2\exp\left(-e^{\imath\psi}\varepsilon\tr r_2\right)\left[D_{k_2r_2}^{(4/\beta)}\left(\imath e^{\imath\psi}\gamma_1\varepsilon\right)\right]^N\frac{\delta(r_2)}{|\Delta_{k_2}(r_2)|^{4/\beta}}\ .
\end{equation}
The constant is
\begin{equation}\label{4.15b}
 C=\displaystyle\left(-\frac{e^{-\imath\psi}}{\gamma_1}\right)^{k_2N}\left(-\frac{\tilde{\gamma}}{2\pi}\right)^{k_1k_2}\left(\frac{2\pi}{\gamma_1}\right)^{k_2}\left(\frac{\pi}{\gamma_1}\right)^{2k_2(k_2-1)/\beta}\frac{G_{Nk_1}^{(\beta)}}{g_{k_2}^{(4/\beta)}}
\end{equation}
with
\begin{equation}\label{c5b}
 \displaystyle g_{k_2}^{(4/\beta)}=\frac{1}{k_2!}\prod\limits_{j=1}^{k_2}\frac{\pi^{2(j-1)/\beta}\Gamma\left(2/\beta\right)}{\Gamma\left(2 j/\beta\right)}\ .
\end{equation}
while the exponent is given by
\begin{equation}\label{c.11}
 \kappa=\frac{N}{\gamma_1}+\frac{\gamma_2-\gamma_1}{2}
\end{equation}
and the differential operator
 \begin{equation}\fl\label{4.15c}
   D_{k_2r_2}^{(4/\beta)}\left(\imath e^{\imath\psi}\gamma_1\varepsilon\right)= \frac{1}{\Delta_{k_2}(r_2)}\det\left[r_{a2}^{N-b}\left(\frac{\partial}{\partial r_{a2}}+(k_2-b)\frac{2}{\beta}\frac{1}{r_{a2}}-e^{\imath\psi}\gamma_1\varepsilon\right)\right]_{1\leq a,b\leq k_2}
 \end{equation}
is the analog to the Sekiguchi differential operator \cite{OkoOls97}. We derived it in \ref{app2}.

The complexity of $D_{k_2r_2}^{(4/\beta)}(\imath e^{\imath\psi}\varepsilon)$ makes Eq. \eref{4.15} cumbersome, a better representation is desirable. To simplify Eq. \eref{4.15}, we need the following statement which is shown in \ref{app3.2}.
\begin{theorem}\label{t1}\ \\
 We consider two functions $F,f:\Herm(4/\beta,k_2)\rightarrow\mathbb{C}$ invariant under the action of $\U^{(4/\beta)}(k_2)$ and Schwartz--functions of the matrix eigenvalues. Let $F$ and $f$ have the relation
 \begin{equation}\label{t1.1}
  F(\rho_2)=f(\rho_2)\det \rho_2^{N/\gamma_1-k}{\rm\ \ for\ all\ }\rho_2\in\Herm(4/\beta,k_2)\ .
 \end{equation}
 Then, we have
 \begin{eqnarray}
   \fl \int\limits_{\mathbb{R}^{k_2}}\int\limits_{\Herm(4/\beta,k_2)}F(r_2){\det}^k r_2|\Delta_{k_2}(r_2)|^{4/\beta} \exp\left(\imath\tr r_2\sigma_2\right){\det}^{N/\gamma_1}\left(e^{-\imath\psi}\sigma_2+\imath\varepsilon\eins_{\tilde{k}}\right)d[\sigma_2]d[r_2]=\nonumber\\
   \fl =  w_1f(0)=\int\limits_{\mathbb{R}^{k_2}}F(r_2)|\Delta_{k_2}(r_2)|^{4/\beta}\left[\frac{w_2\exp\left(\varepsilon e^{\imath\psi}\tr r_2\right)}{|\Delta_{k_2}(r_2)|^{4/\beta}}\prod\limits_{j=1}^{k_2}\left(\frac{\partial}{\partial r_{j2}}\right)^{N-k_1}\delta(r_{j2})\right]d[r_2]\label{t1.2}
 \end{eqnarray}
 where the constants are
 \begin{eqnarray}
   \fl w_1 =  \left(\frac{2\pi}{\gamma_1}\right)^{k_2}\left(\frac{\pi}{\gamma_1}\right)^{2k_2(k_2-1)/\beta}\frac{\left(\imath^N e^{-\imath\psi N}\right)^{k_2}}{g_{k_2}^{(4/\beta)}}\prod_{b=1}^{k_2}\prod\limits_{a=1}^N\left(\frac{a}{\gamma_1}+\frac{b-1}{\gamma_2}\right)\label{c.6b}\\
   \fl w_2 = \frac{(-1)^{k_1k_2}}{g_{k_2}^{(4/\beta)}}\left(\frac{2\pi}{\gamma_1}\right)^{k_2}\left(\frac{\pi}{\gamma_1}\right)^{2k_2(k_2-1)/\beta}\left[\frac{(-\imath)^N e^{-\imath\psi N}}{\left(N-k_1\right)!\gamma_1^N}\right]^{k_2} \prod_{j=0}^{k_2-1}\frac{\Gamma\left(N+1+2j/\beta\right)}{\Gamma\left(1+2j/\beta\right)}\ .\label{c.6}
 \end{eqnarray}
\end{theorem}

This statement yields for the supersymmetric Ingham--Siegel integral
\begin{equation}\label{4.16}
  I_k^{(\beta,N)}(\rho)=\displaystyle W\Theta(r_1)\frac{{\det}^{\kappa} r_1}{|\Delta_{k_2}(r_2)|^{4/\beta}}\prod\limits_{j=1}^{k_2}\left(\frac{\partial}{\partial r_{j2}}\right)^{N-k_1}\delta(r_{j2})
\end{equation}
where the constant reads
\begin{eqnarray}
  W & = & \left(\frac{\tilde{\gamma}}{2\pi}\right)^{k_1k_2}\left(\frac{2\pi}{\gamma_1}\right)^{k_2}\left(\frac{\pi}{\gamma_1}\right)^{2k_2(k_2-1)/\beta}\left[\frac{\left(-e^{-\imath\psi}\right)^N}{\left(N-k_1\right)!\gamma_1^N}\right]^{k_2}\times\nonumber\\
  & \times &  \frac{G_{Nk_1}^{(\beta)}}{g_{k_2}^{(4/\beta)}}\prod_{j=0}^{k_2-1}\frac{\Gamma\left(N+1+2j/\beta\right)}{\Gamma\left(1+2j/\beta\right)}\ .\label{c.12}
\end{eqnarray}
We further simplify this formula for $\beta=1$ and $\beta=2$. The powers of the Vandermonde--determinant  $\Delta_{k_2}^{4/\beta}(r_2)$ are polynomials of degree $k_2\times2(k_2-1)/\beta$. The single power of one eigenvalue derivative must be $2(k_2-1)/\beta$ if we substitute these terms in Eq. \eref{4.16} by partial derivatives of the eigenvalues, for details see \ref{app3.2}. Hence, this power is a half-integer for $\beta=4$. Also, $\Delta_{k_2}(r_2)$ has no symmetric term where all eigenvalues have the same power. Therefore, we can not simplify the quaternionic case in the same manner.

We use the identities
\begin{eqnarray}
 \prod\limits_{j=1}^n\frac{\partial^{n-1}}{\partial x_j^{n-1}}\Delta_n^2(x) & = & (-1)^{n(n-1)/2}n!\left[(n-1)!\right]^n\ ,\label{4.17}\\
 \prod\limits_{j=1}^n\frac{\partial^{2(n-1)}}{\partial x_j^{2(n-1)}}\Delta_n^4(x) & = & n!\left[(2n-2)!\right]^n\prod\limits_{j=0}^{n-1}(2j+1)\label{4.18}
\end{eqnarray}
and find
\begin{eqnarray}
  I_k^{(1,N)}(\rho) & = & 2^{-k(k-2)}\left[\frac{2\pi e^{-\imath\psi N}}{(N-2)!}\right]^k\times\nonumber\\
  & \times & \Theta(r_1)\det r_1^{(N-1)/2}\prod\limits_{j=1}^{k}\left(-\frac{\partial}{\partial r_{j2}}\right)^{N-2}\delta(r_{j2})\label{4.19}
\end{eqnarray}
and
\begin{eqnarray}
  I_k^{(2,N)}(\rho) & = & (-1)^{k(k+1)/2}2^{-k(k-1)}\left[\frac{2\pi e^{-\imath\psi N}}{(N-1)!}\right]^k\times\nonumber\\
  & \times & \Theta(r_1)\det r_1^{N}\prod\limits_{j=1}^{k}\left(-\frac{\partial}{\partial r_{j2}}\right)^{N-1}\delta(r_{j2})\ .\label{4.20}
\end{eqnarray}
For $\beta=4$, we summarize the constants and have
\begin{eqnarray}
  I_k^{(4,N)}(\rho) & = & 2^{-k(k-2)}\left[\frac{2\pi e^{-\imath\psi N}}{(N-k)!}\right]^{2k}\times\nonumber\\
  & \times & \Theta(r_1)\det r_1^{N+1/2}\frac{4^kk!}{\pi^k|\Delta_{2k}(r_2)|}\prod\limits_{j=1}^{2k}\left(-\frac{\partial}{\partial r_{j2}}\right)^{N-k}\delta(r_{j2})\ .\label{4.21}
\end{eqnarray}
These distributions are true for superfunctions whose Fermion--Fermion block dependence is as in Eq. \eref{t1.1}. Eqs. \eref{4.19} and \eref{4.20} can be extended to distributions on arbitrary Schwartz--functions which is not the case for Eq. \eref{4.21}. The constants in Eqs. \eref{4.19} and \eref{4.20} must be the same due to the independence of the test--function.
\begin{theorem}\label{t5}\ \\
 Equations \eref{4.19} and \eref{4.20} are true for rotation invariant superfunctions $\Phi_0$ which are Schwartz--functions in the Fermion--Fermion block entries along the Wick--rotated real axis.
\end{theorem}
We derive this statement in \ref{app3.3}.

Indeed, the Eq. \eref{4.20} is the same as the formula for the supersymmetric Ingham--Siegel integral for $\beta=2$ in Ref.  \cite{Guh06}. Comparing both results, the different definitions of the measures have to be taken into account. We also see the similarity to the superbosonization formula \cite{EfeKog04,EST04,LSZ07,Som07,BEKYZ07,BasAke07} for $\beta\in\{1,2\}$. One can replace the partial derivative in Eq. \eref{4.19} and \eref{4.20} by contour integrals if the characteristic function $\Phi_0$ is analytic. However for $\beta=4$, more effort is needed. For our purposes, Eqs. \eref{4.15} and \eref{4.21} are sufficient for the quaternionic case. In the unitary case, the equivalence of Eq. \eref{4.20} with the superbosonization formula was confirmed with help of Cauchy integrals by Basile and Akemann. \cite{BasAke07}

\section{Final representation of the generating function and its independence of the choice for $\Phi_0$}\label{sec5}

In Sec. \ref{sec5.1}, we present the generating function as a supersymmetric integral over eigenvalues and introduce the supersymmetric Bessel--functions. In Sec. \ref{sec5.2}, we revisit the unitary case and point out certain properties of the generating function. Some of these properties, independence of the Wick--rotation and the choice of $\Phi_0$, are also proven for the orthogonal and unitary--symplectic case in Sec. \ref{sec5.3}.

\subsection{Eigenvalue integral representation}\label{sec5.1}

The next step of the calculation of the generating function $Z_k(x^-+J)$ is the integration over the supergroup. The function $\Phi_0(\rho)I_k^{(\beta,N)}(\rho)$ is invariant under the action of $\U^{(\beta)}(k_1/k_2)$.

We define the supermatrix Bessel--function
\begin{equation}\label{5.2}
 \varphi_{k_1k_2}^{(\beta)}(s,r)=\int\limits_{\U^{(\beta)}(k_1/k_2)}\exp\left(\Str sUrU^\dagger\right)d\mu(U)
\end{equation}
as in Refs.  \cite{GuhKoh02b,KKG08}. We choose the normalization
\begin{eqnarray}
  & & \int\limits_{\Sigma^0_\psi(\beta,k)} f(\sigma)\exp\left(\Str\sigma x\right)d[e^{-\imath\psi/2}\eta]d[e^{\imath\psi}\sigma_2]d[\sigma_1]=\nonumber\\
  & = & \int\limits_{\mathbb{R}^{k_1}}\int\limits_{\mathbb{R}^{k_2}}f(s)\varphi_{k_1k_2}^{(\beta)}(s,x)\left|B_{k}^{(\beta)}(s_1,e^{\imath\psi}s_2)\right|d[e^{\imath\psi}s_2]d[s_1]+{\rm b.t.}\label{5.3}
\end{eqnarray}
which holds for every rotation invariant function $f$. This normalization agrees with Refs.  \cite{Guh96,Guh96b,GuhKoh02b,Guh06,KKG08}. The boundary terms (${\rm b.t.}$) referred to as Efetov--Wegner terms \cite{Guh91,Guh93,BasAke07} appear upon changing the integration variables \cite{Rot87} or, equivalently, upon partial integration \cite{KKG08}. The Berezinian is
\begin{equation}\label{5.6}
 B_{k}^{(\beta)}(s_1,e^{\imath\psi}s_2)=\displaystyle\frac{\Delta_{k_1}^\beta(s_1)\Delta_{k_2}^{4/\beta}(e^{\imath\psi}s_2)}{V_{k}^2(s_1,e^{\imath\psi}s_2)}
\end{equation}
where $V_{k}(s_1,e^{\imath\psi}s_2)=\prod\limits_{n=1}^{k_1}\prod\limits_{m=1}^{k_2}\left(s_{n1}-e^{\imath\psi}s_{m2}\right)$ mixes bosonic and fermionic eigenvalues. These Berezinians have a determinantal structure
\begin{equation}\fl\label{5.7}
 B_k^{(\beta)}(s_1,e^{\imath\psi}s_2)=\left\{\begin{array}{ll}
                                                      \displaystyle\det\left[\frac{1}{s_{a1}-e^{\imath\psi}s_{b2}}\ ,\ \frac{1}{(s_{a1}-e^{\imath\psi}s_{b2})^2}\right]\underset{1\leq b\leq k}{\underset{1\leq a\leq 2k}{ }} & ,\ \beta=1\\
                                                      \displaystyle{\det}^{2}\left[\frac{1}{s_{a1}-e^{\imath\psi}s_{b2}}\right]_{1\le a,b\le k} & ,\ \beta=2\\
                                                      \displaystyle B_k^{(1)}(e^{\imath\psi}s_2,s_1) & ,\ \beta=4
                                                     \end{array}\right.\ .
\end{equation}
For $\beta=2$ this formula was derived in Ref.  \cite{Guh91}. The other cases are derived in \ref{app4}. We notice that this determinantal structure is similar to the determinantal structure of the ordinary Vandermonde--determinant raised to the powers $2$ and $4$. This structure was explicitly used \cite{Meh67} to calculate the $k$--point correlation function of the GUE and the GSE.

We find for the generating function
\begin{eqnarray}\fl
  Z_k(x^-+J) & = & 2^{2k(k-\tilde{\gamma})}e^{\imath\psi k_1}\int\limits_{\mathbb{R}^{k_1}}\int\limits_{\mathbb{R}^{k_2}}\Phi_0(r)I_k^{(\beta,N)}(r)\times\nonumber\\
  & \times & \varphi_{k_1k_2}^{(\beta)}(-\imath r,x^-+J)\left| B_{k}^{(\beta)}(r_1,e^{\imath\psi}r_2)\right|d[r_2]d[r_1]+{\rm b.t.}\label{5.8}\ .
\end{eqnarray}
The normalization of $Z_k$ is guaranteed by the Efetov--Wegner terms. When setting $(k-l)$ with $l<k$ of the source variables $J_p$ to zero then we have
\begin{equation}\label{5.4}
 \left.Z_k(x^-+J)\right|_{J_l=\ldots=J_k=0}=Z_{l-1}(\tilde{x}^-+\widetilde{J})\ ,
\end{equation}
$\tilde{x}=\diag(x_1,\ldots,x_{l-1}),\ \widetilde{J}=\diag(J_1,\ldots,J_{l-1})$, by the integration theorems in Ref.  \cite{Efe83,Weg83,Con88,ConGro89,Efe97,KKG08}. This agrees with the definition \eref{2.8}.

\subsection{The unitary case revisited}\label{sec5.2}

To make contact with the discussion in Ref.  \cite{Guh06}, we revisit the unitary case using the insight developed here.

For a further calculation we need the explicit structure of the supersymmetric matrix Bessel--functions. However, the knowledge of these functions is limited. Only for certain $\beta$ and $k$ we know the exact structure. In particular for $\beta=2$ the supermatrix Bessel--function was first calculated in Ref.  \cite{Guh91,Guh96} with help of the heat equation. Recently, this function was re-derived by integrating the Grassmann variables in Cartesian coordinates \cite{KKG08},
\begin{eqnarray}\fl
 \varphi_{kk}^{(2)}(-\imath r,x^-+J)=\displaystyle\frac{\imath^k\exp\left(-\varepsilon\Str r\right)}{2^{k^2}\pi^k}\times\nonumber\\
\fl\times\frac{\det\left[\exp\left(-\imath r_{m1}(x_{n}-J_{n})\right)\right]_{1\leq m,n\leq k}\det\left[\exp\left(\imath e^{\imath\psi}r_{m2}(x_{n}+J_{n})\right)\right]_{1\leq m,n\leq k}}{\sqrt{B_{k}^{(2)}(r_1,e^{\imath\psi}r_2)B_{k}^{(2)}\left(x-J,x+J\right)}}\label{5.12}
\end{eqnarray}
with $x\pm J=\diag(x_1\pm J_1,\ldots,x_k\pm J_k)$ and the positive square root of the Berezinian
\begin{equation}\fl\label{5.13}
 \displaystyle\sqrt{B_{k}^{(2,2)}(r_1,e^{\imath\psi}r_2)}=\displaystyle\det\left[\frac{1}{r_{a1}-e^{\imath\psi}r_{b2}}\right]_{1\le a,b\le k}=(-1)^{k(k-1)/2}\frac{\Delta_{k}(s_1)\Delta_{k}(e^{\imath\psi}s_2)}{V_{k}(s_1,e^{\imath\psi}s_2)}\ .
\end{equation}
Due to the structure of $\varphi_{kk}^{(2)}$ and $B_{k}^{(2)}$, we write the generating function for $\beta=2$ as an integral over $\Phi_0$ times a determinant \cite{Guh06}
\begin{eqnarray}\fl
  Z_k(x^-+J) & = &(-1)^{k(k+1)/2}\displaystyle{\det}^{-1}\left[\frac{1}{x_{a}-x_{b}-J_a-J_b}\right]_{1\le a,b\le k}\int\limits_{\mathbb{R}^{k}}\int\limits_{\mathbb{R}^{k}}\Phi_0(r)\times\nonumber\\
  & \times & \det\left[\mathfrak{F}_N(\tilde{r}_{mn},\tilde{x}_{mn})\Theta(r_{m1})\exp\left(-\varepsilon\Str\tilde{r}_{mn}\right)\right]_{1\leq m,n\leq k}d[r_2]d[r_1]+{\rm b.t.}\label{5.14}
\end{eqnarray}
where $\tilde{r}_{mn}=\diag\left(r_{m1},e^{\imath\psi}r_{n2}\right)$, $\tilde{x}_{mn}=\diag\left(x_{m}-J_m,x_{n}+J_n\right)$ and
\begin{equation}\label{5.15}
\mathfrak{F}_N(\tilde{r}_{mn},\tilde{x}_{mn})=\frac{\imath r_{m1}^N\exp\left(-\imath\Str\tilde{r}_{mn}\tilde{x}_{mn}\right)}{(N-1)!(r_{m1}-e^{\imath\psi}r_{n2})}\left(-e^{-\imath\psi}\frac{\partial}{\partial r_{n2}}\right)^{N-1}\delta(r_{n2})\ .
\end{equation}
Then, the modified $k$--point correlation function is
\begin{eqnarray}\fl
 \qquad\quad\widehat{R}_k(x^-) & = & \int\limits_{\mathbb{R}^{k}}\int\limits_{\mathbb{R}^{k}}\Phi_0(r)\times\nonumber\\
 & \times & \det\left[\mathfrak{F}_N(\tilde{r}_{mn},x_{mn})\Theta(r_{m1})\exp\left(-\varepsilon\Str\tilde{r}_{mn}\right)\right]_{1\leq m,n\leq k}d[r_2]d[r_1]+{\rm b.t.}\label{5.16}
\end{eqnarray}
and the $k$--point correlation function is
\begin{equation}\label{5.17}
 R_k(x)=\int\limits_{\mathbb{R}^{k}}\int\limits_{\mathbb{R}^{k}}\Phi_0(r)\det\left[\frac{\mathfrak{F}_N(\tilde{r}_{mn},x_{mn})}{2\pi\imath}\right]_{1\leq m,n\leq k}d[r_2]d[r_1]+{\rm b.t.}\ .
\end{equation}
We defined $x_{mn}=\diag(x_m,x_n)$. The boundary terms comprise the lower correlation functions. The $k$--point correlation function for $\beta=2$ is a determinant of the fundamental function
\begin{equation}\label{5.18}
 R^{({\rm fund})}(x_m,x_n)=\int\limits_{\mathbb{R}}\int\limits_{\mathbb{R}}\Phi_0(r)\frac{\mathfrak{F}_N(r,x_{mn})}{2\pi\imath}dr_2dr_1
\end{equation}
if there is one characteristic function $\mathcal{F}P_0$ with a supersymmetric extension $\Phi_0$ factorizing for diagonal supermatrices,
\begin{equation}\fl\label{5.19}
 \Phi_0(r)=\Sdet\diag\left[\widehat\Phi_0(r_{11}),\ldots,\widehat\Phi_0(r_{k1}),\widehat\Phi_0\left(e^{\imath\psi}r_{12}\right),\ldots,\widehat\Phi_0\left(e^{\imath\psi}r_{k2}\right)\right]\ ,
\end{equation}
with $\widehat\Phi_0:\mathbb{C}\rightarrow\mathbb{C}$. For example, the shifted Gaussian ensemble in App. F of Ref.  \cite{Guh06} is of such a type.

In Eq. \eref{5.18} we notice that this expression is independent of the generalized Wick--rotation. Every derivative of the fermionic eigenvalue $r_2$ contains the inverse Wick--rotation as a prefactor. Moreover, the Wick--rotation in the functions are only prefactors of $r_2$. Thus, an integration over the fermionic eigenvalues $r_2$ in Eq. \eref{5.16} cancels the Wick--rotation out by using the Dirac--distribution. Also, this integration shows that every representation of the characteristic function gives the same result, see Theorem \ref{t2} in the next subsection. However, the determinantal structure with the fundamental function in Eq. \eref{5.18} depends on a special choice of $\Phi_0$.

\subsection{Independence statement}\label{sec5.3}

For $\beta=1$ and $\beta=4$ we do not know the ordinary matrix Bessel--function explicitly. Hence, we can not give such a compact expression as in the case $\beta=2$. On the other hand, we can derive the independence of the Wick--rotation and of the $\Phi_0$ choice of the generating function.
\begin{theorem}\label{t2}\ \\
 The generating function $Z_k$ is independent of the Wick--rotation and of the choice of the characteristic functions supersymmetric extension $\Phi_0$ corresponding to a certain matrix ensemble $(P,\Herm(\beta,N))$.
\end{theorem}
\textbf{Derivation:}\\
We split the derivation in two parts. The first part regards the Wick--rotation and the second part yields the independence of the choice of $\Phi_0$.

Due to the normalization of the supermatrix Bessel--function \eref{5.3}, $\varphi_{k_1k_2}^{(\beta)}(-\imath r,x^-+J)$ only depends on $e^{\imath\psi}r_2$. The same is true for $\Phi_0$. Due to the property
\begin{equation}\label{tp2.2b}
 D_{k_2r_2}^{(4/\beta)}\left(\imath e^{\imath\psi}\gamma_1\varepsilon\right)=e^{\imath k_2\psi}D_{k_2,e^{\imath\psi}r_2}^{(4/\beta)}\left(\imath \gamma_1\varepsilon\right)\ ,
\end{equation}
the Ingham--Siegel integral in the form \eref{4.15} times the phase $e^{\imath(k_1-k_2)\psi}$ only depends on $e^{\imath\psi}r_2$ and $e^{-\imath\psi}\partial/\partial r_2$. The additional phase comes from the $\rho$--integration. Thus, we see the independence of the Wick--rotation because of the same reason as in the $\beta=2$ case.

Let $\Phi_0$ and $\Phi_1$ be two different supersymmetric extensions of the characteristic function $\mathcal{F}P$. Then these two superfunctions only depend on the invariants $\{\Str\sigma^{m_j}\}_{1\leq j\leq l_0}$ and $\{\Str\sigma^{n_j}\}_{1\leq j\leq l_1}$, $m_j,n_j,l_0,l_1\in\mathbb{N}$. We consider $\Phi_0$ and $\Phi_1$ as functions of $\mathbb{C}^{l_0}\rightarrow\mathbb{C}$ and $\mathbb{C}^{l_1}\rightarrow\mathbb{C}$, respectively. Defining the function
\begin{equation}\label{tp2.3}
 \Delta\Phi(x_1,\ldots,x_{M})=\Phi_0(x_{m_1},\ldots,x_{m_{l_0}})-\Phi_1(x_{n_1},\ldots,x_{n_{l_1}}),
\end{equation}
where $M={\rm max}\{m_a,n_b\}$, we notice with the discussion in Sec. \ref{sec3.5} that
\begin{equation}\label{tp2.4}
 \Delta\Phi(x_1,\ldots,x_{M})|_{x_j=\tr H^j}=0
\end{equation}
for every hermitian matrix $H$. However, there could be a symmetric supermatrix $\sigma$ with
\begin{equation}\label{tp2.5}
 \Delta\Phi(x_1,\ldots,x_{M})|_{x_j=\Str \sigma^j}\neq0.
\end{equation}
With the differential operator
\begin{equation}\label{tp2.6}
 \mathfrak{D}_r=\left[D_{k_2r_2}^{(4/\beta)}\left(\imath e^{\imath\psi}\gamma_1\varepsilon\right)\right]^{N-k_1}\frac{\varphi_{k_1k_2}^{(\beta)}(-\imath r,x^-+J)}{V_k(r_1,e^{\imath\psi}r_2)},
\end{equation}
we consider the difference of the generating functions
\begin{eqnarray}
 \fl\Delta Z_k(x^-+J) & = & Z_k(x^-+J)|_{\Phi_0}-Z_k(x^-+J)|_{\Phi_1}=\nonumber\\
 \fl& = & \int_{\mathbb{R}^{k_1}}|\Delta_{k_2}(r_1)|^{\beta}{\det}^\kappa r_1\Theta(r_1)\left.\mathfrak{D}_r\Delta\Phi(x)|_{x_j=\Str r^j}\right|_{r_2=0}d[r_1]\label{tp2.7}
\end{eqnarray}
Here, we omit the Efetov--Wegner terms. The differential operator is invariant under the action of the permutation group $S(k_2)$ on the fermionic block $\Herm(4/\beta,k_2)$. Hence, we find
\begin{eqnarray}
 \fl\left.\mathfrak{D}_r\Delta\Phi(x)|_{x_j=\Str r^j}\right|_{r_2=0} & = & \left.\underset{|a|\leq k_2(N-k_1)}{\sum\limits_{a\in\{0,\ldots,N-k_1\}^M}}d_a(r)\prod\limits_{j=1}^M\frac{\partial^{a_j}}{\partial x_j^{a_j}}\Delta\Phi(x)|_{x_j=\Str r^j}\right|_{r_2=0}=\nonumber\\
 \fl & = & \underset{|a|\leq k_2(N-k_1)}{\sum\limits_{a\in\{0,\ldots,N-k_1\}^M}}d_a(r_1)\prod\limits_{j=1}^M\frac{\partial^{a_j}}{\partial x_j^{a_j}}\Delta\Phi(x)|_{x_j=\tr r^j}=\nonumber\\
 \fl & = & 0\label{tp2.8},
\end{eqnarray}
where $d_a$ are certain symmetric functions depending on the eigenvalues $r$.  At $r_2=0$ these functions are well-defined since the supermatrix Bessel--functions and the term $V_k^{-1}(r_1,e^{\imath\psi}r_2)$ are $C^{\infty}$ at this point. Thus, we find that
\begin{equation}\label{tp2.9}
 \Delta Z_k(x^-+J)=0.
\end{equation}
This means that the generating function is independent of the supersymmetric extension of the characteristic function.\hfill$\square$

\section{One--point and higher order correlation functions}\label{sec6}

We need an explicit expression or some properties of the supermatrix Bessel--function to simplify the integral for the generating function. For $k=1$ we know the supermatrix Bessel--functions for all $\beta$. The simplest case is $\beta=2$ where we take the formula \eref{5.17} with $k=1$ and obtain
\begin{equation}\label{6.1}
 R_1(x)=R^{({\rm fund})}(x,x)=\int\limits_{\mathbb{R}}\int\limits_{\mathbb{R}}\Phi_0(r)\frac{\mathfrak{F}_N\left(r,x\eins_2\right)}{2\pi\imath}dr_2dr_1\ .
\end{equation}
Since the Efetov--Wegner term in the generating function is just unity there are no boundary terms in the level density. For $\beta\in\{1,4\}$ we use the supermatrix Bessel--function \cite{GuhKoh02b,BreHik03,KKG08}
\begin{eqnarray}
  \varphi_{21}^{(1)}(-\imath r,x^-+J) & = & \frac{-2J}{\pi}\exp\left[-\imath\Str r(x^-+J)\right]\times\nonumber\\
 &\times & \left[\imath\Str r+J\left(r_{11}-e^{\imath\psi}r_{2}\right)\left(r_{21}-e^{\imath\psi}r_{2}\right)\right]\ .\label{6.2}
\end{eqnarray}
We find
\begin{eqnarray}\fl
  \widehat{R}_1(x^-) = \displaystyle-\imath \int\limits_{\mathbb{R}^2}\int\limits_{\mathbb{R}}\Phi_0(r)\det r_1^{(N-1)/2}\Str r\frac{|r_{11}-r_{21}|}{(r_{11}-e^{\imath\psi}r_2)^2(r_{21}-e^{\imath\psi}r_2)^2}\times\nonumber\\
  \times \displaystyle \exp\left(-\imath x^-\Str r\right)\Theta(r_1)\frac{1}{(N-2)!}\left(-e^{-\imath\psi}\frac{\partial}{\partial r_2}\right)^{N-2}\delta(r_2)d[r_1]dr_2\label{6.3}
\end{eqnarray}
for $\beta=1$ and
\begin{eqnarray}\fl
  \widehat{R}_1(x^-) = \displaystyle-4\imath\int\limits_{\mathbb{R}}\int\limits_{\mathbb{R}^2}\Phi_0(r) r_1^{2N+1}\Str r\frac{e^{\imath\psi}r_{12}-e^{\imath\psi}r_{22}}{(r_1-e^{\imath\psi}r_{12})^2(r_1-e^{\imath\psi}r_{22})^2}\times\nonumber\\
 \times \exp\left(-\imath x^-\Str r\right)\Theta(r_1)\frac{\det e^{\imath\psi}r_2}{(2N+1)!}\left(4e^{-2\imath\psi}D_{2,r_2}^{(1)}\right)^N\frac{\delta(r_{12})\delta(r_{22})}{e^{\imath\psi}r_{12}-e^{\imath\psi}r_{22}}d[r_2]dr_1\label{6.4}
\end{eqnarray}
for $\beta=4$. The differential operator has the explicit form
\begin{equation}\label{6.4b}
 D_{2,r_2}^{(1)}=\frac{\partial^2}{\partial r_{12}\partial r_{22}}-\frac{1}{2}\frac{1}{r_{12}-r_{22}}\left(\frac{\partial}{\partial r_{12}}-\frac{\partial}{\partial r_{22}}\right)\ .
\end{equation}
For the level density we have
\begin{eqnarray}\fl
  R_1(x)  =  \displaystyle -\frac{1}{2\pi}\int\limits_{\mathbb{R}^2}\int\limits_{\mathbb{R}}\Phi_0(r)\det r_1^{(N-1)/2}\exp\left(-\imath x\Str r\right)\Str r\frac{|r_{11}-r_{21}|}{(r_{11}-e^{\imath\psi}r_2)^2(r_{21}-e^{\imath\psi}r_2)^2}\times\nonumber\\
   \times  \displaystyle \left(\Theta(r_1)+\Theta(-r_1)\right)\frac{1}{(N-2)!}\left(-e^{-\imath\psi}\frac{\partial}{\partial r_2}\right)^{N-2}\delta(r_2)d[r_1]dr_2\label{6.5}
\end{eqnarray}
for $\beta=1$ and
\begin{eqnarray}\fl
  R_1(x)  =  \displaystyle-\frac{2}{\pi}\int\limits_{\mathbb{R}}\int\limits_{\mathbb{R}^2}\Phi_0(r) r_1^{2N+1}\exp\left(-\imath x\Str r\right)\Str r\frac{e^{\imath\psi}r_{12}-e^{\imath\psi}r_{22}}{(r_1-e^{\imath\psi}r_{12})^2(r_1-e^{\imath\psi}r_{22})^2}\times\nonumber\\
  \times  \frac{\det e^{\imath\psi}r_2}{(2N+1)!}\left(4e^{-2\imath\psi}D_{2,r_2}^{(1)}\right)^N\frac{\delta(r_{12})\delta(r_{22})}{e^{\imath\psi}r_{12}-e^{\imath\psi}r_{22}}d[r_2]dr_1\label{6.6}
\end{eqnarray}
for $\beta=4$. The equations \eref{6.4} to \eref{6.6} comprise all level--densities for arbitrary matrix ensembles invariant under orthogonal and unitary--symplectic rotations. As probability densities which do not factorize are included, these results considerably extend those obtained by orthogonal polynomials.

For higher order correlation functions we use the definition \eref{2.4} and the definition of the matrix Green's function. With help of the quantities $L=\diag(L_1,\ldots,L_k)\in\{\pm 1\}^k$ and $\widehat{L}=L\otimes\eins_{2\tilde{\gamma}}$, this yields
\begin{eqnarray}\fl
  R_k(x)  =  \displaystyle 2^{2k(k-\tilde{\gamma})}\int\limits_{\mathbb{R}^{k_1}}\int\limits_{\mathbb{R}^{k_2}}\Phi_0(r)\underset{\epsilon\searrow 0}{\lim}\sum\limits_{L\in\{\pm 1\}^k} \prod\limits_{j=1}^kL_j\ \frac{I_k^{(\beta,N)}\left(\widehat{L}r\right)\exp\left(-\varepsilon\Str\widehat{L}r\right)}{\left(2\pi\imath e^{-\imath\psi\gamma_1}\right)^k}\times\nonumber\\
   \fl\times \displaystyle\left.\left(\prod\limits_{j=1}^k-\frac{1}{2}\frac{\partial}{\partial J_j}\right)\varphi_{k_1k_2}^{(\beta)}(-\imath r,x^{(0)}+J)\right|_{J=0}\left| B_{k}^{(\beta)}(r_1,e^{\imath\psi}r_2)\right|d[r_2]d[r_1]+{\rm b.t.}\label{6.7}
\end{eqnarray}
for analytic correlation functions. We extend this formula to all rotation invariant ensembles by the universality of the integral kernel. First, we make a limit of a uniformly convergent series of Schwartz--functions analytic in the real components of its entries to every arbitrary Schwartz--function describing a matrix ensemble. The Schwartz--functions are dense in a weak sense in the sets of Lebesgue--integrable Functions $L^p$ and the tempered distributions. Thus, we integrate Eq.~\eref{6.7} with an arbitrary Schwartz--function on $\mathbb{R}^k$ and take the limit of a series of Schwartz--functions describing the ensembles to a tempered distribution which completes the extension.

\section{Remarks and conclusions}\label{sec7}

We extended the method of the generalized Hubbard--Stratonovich transformation to arbitrary orthogonally and unitary--symplectically invariant random matrix ensembles. Due to a duality between ordinary and supersymmetric matrix spaces, the integral for the $k$--point correlation function is over a superspace. This integral was reduced to an eigenvalue integral for all probability densities, including those which do not factorize. The results are in terms of the characteristic function. Thus, the characteristic function has to be calculated for the ensemble in question. Since the matrix Bessel--functions of the ordinary orthogonal and unitary--symplectic group \cite{GuhKoh02,GuhKoh02b,BerEyn08} and, thus, the supermatrix Bessel--functions of $\UOSp(2k/2k)$ are not known explicitly beyond $k=1$, we can not further simplify our results. However, we found the previously unknown determinantal structure of the Berezinian of $\UOSp(2k/2k)$.

Up to the restriction $N\geq k_1$, formula \eref{6.7} is exact for every $k$, $N$ and rotation invariant ensemble. Thus, it can serve not only as starting point for universality considerations \cite{HacWei95}, but for all other studies.

The expressions for the supersymmetric Ingham--Siegel integrals \eref{4.19}, \eref{4.20} and \eref{4.21} confirm the equivalence of the superbosonization formula \cite{BEKYZ07,Som07,LSZ07} with our derivation. A work for a proof of this equivalence for all $\beta$'s is in progress. The comparison of the superbosonization formula \cite{LSZ07,Som07} with Eq. \eref{4.1} shows that the crucial difference lies in the integration domain. However, the Dirac--distribution and the partial derivatives in the fermionic part imply a representation as a contour integral which is equivalent to the compact space used in the superbosonization formula.

\section*{Acknowledgements}
We thank H. Kohler for clarifying remarks on relation between the
ordinary matrix Bessel--functions and the Jack--polynomials as well as
on the Sekiguchi differential operators. We are also grateful to S. Mandt,
H.-J. Sommers and M.R.  Zirnbauer for fruitful discussions. A big thank you goes to P. Heinzner and E. Vishnyakova for helpful advice on the Paley--Wiener theorem. We thank the referee for helpful remarks. We
acknowledge financial support from the Deutsche Forschungsgemeinschaft
within Sonderforschungsbereich Transregio 12 ``Symmetries and Universality in Mesoscopic Systems''
(M.K. and T.G.) and from Det Svenska Vetenskapsr\aa det (J.G.).

\appendix

\section{Circularity of the supertrace for rectangular supermatrices}\label{app1}

The circularity for rectangular matrices of pure commuting entries or anticommuting entries was derived by Berezin \cite{Ber87}. Since we have not found the general theorem for arbitrary rectangular supermatrices, we give the trivial statement.
\begin{corollary}\label{at2}\ \\
 Let the matrices $V_1$ and $V_2$ be the same as in Eq. \eref{3.23}. Then, we have
 \begin{equation}\label{at2.1}
  \Str V_1V_2=\Str V_2V_1
 \end{equation}
\end{corollary}
\textbf{Derivation:}\\
We recall the circularity of the trace for rectangular matrices of commuting elements $\tr A_1A_2=\tr A_2A_1$ and its anticommuting analogue $\tr B_1B_2=-\tr B_2B_1$ which has been proven by Berezin \cite{Ber87}. We make the simple calculation
\begin{eqnarray}
  \Str V_1V_2 & = & \tr A_1A_2+\tr B_1C_2-\tr C_1B_2-\tr D_1D_2\nonumber\\
  & = & \tr A_2A_1-\tr C_2B_1+\tr B_2C_1-\tr D_2D_1\nonumber\\
  & = & \Str V_2V_1\label{a2.1}
\end{eqnarray}
\hfill$\square$

For our purposes we must prove
\begin{equation}\label{a2.2}
 \tr(V^\dagger V)^m=\Str(VV^\dagger)^m\ .
\end{equation}
We define $V_1=V^\dagger$ and $V_2=(VV^\dagger)^{m-1}V$ and get $a=2k$, $b=2k$, $c=\gamma_2N$ and $d=0$. Applying corollary \ref{at2} and reminding that $\tr A=\Str A$ for a matrix of commuting elements and identification with the Boson--Boson block, we have the desired result \eref{a2.2}.

\section{A matrix--Bessel version of the Sekiguchi differential operator}\label{app2}

We derive a version for the Sekiguchi differential operator for the ordinary matrix Bessel--functions $\varphi_{N}^{(\beta)}(y,x)$ on the connection between the Jack--polynomials and the ordinary matrix Bessel--functions.

The Sekiguchi differential operator is defined as \cite{OkoOls97}
\begin{eqnarray}\fl
  D_{Nz}(u,\beta)  =  \Delta_N^{-1}(z)\det\left[z_a^{N-b}\left(z_a\frac{\partial}{\partial z_a}+(N-b)\frac{\beta}{2}+u\right)\right]_{1\leq a,b\leq N}=\nonumber\\
   \fl=  \Delta_N^{-1}(z)\det\left[\frac{\beta}{2}\left(z_a\frac{\partial}{\partial z_a}+u\right)z_a^{N-b}+\left(1-\frac{\beta}{2}\right)z_a^{N-b}\left(z_a\frac{\partial}{\partial z_a}+u\right)\right]_{1\leq a,b\leq N}\ .\label{a3.1}
\end{eqnarray}
Here, $u$ is a boost and the expansion parameter to generate the elementary polynomials in the Cherednik operators, for more explicit information see Ref.  \cite{FJMM02}. Let $J_{N}^{(\beta)}(n,z)$ the Jack--polynomial with the partition $n_1\geq\ldots\geq n_N$ and the standard parameter $\alpha=\frac{2}{\beta}$ in Macdonald's \cite{Mac95} notation. The Jack--polynomials are eigenfunctions with respect to $D_{Nz}(u,\beta)$
\begin{equation}\label{a3.2}
 D_{Nz}(u,\beta)J_{N}^{(\beta)}(n,z)=\prod\limits_{a=1}^N\left[n_a+(N-a)\frac{\beta}{2}+u\right]J_{N}^{(\beta)}(n,z)\ .
\end{equation}
The aim is to find a similar differential operator for the ordinary matrix Bessel--function $\varphi_{N}^{(\beta)}(y,x)$ such that
\begin{eqnarray}
 D_{Nx}^{(\beta)}(B)\varphi_{N}^{(\beta)}\left(\frac{y}{\gamma_2},x\right) & = & \prod\limits_{a=1}^N\imath\left(y_a+B\right)\varphi_{N}^{(\beta)}\left(\frac{y}{\gamma_2},x\right)=\nonumber\\
 & = & {\det}^{1/\gamma_2} \imath(y+ B\eins_{\gamma_2N})\varphi_{N}^{(\beta)}\left(\frac{y}{\gamma_2},x\right).\label{a3.3}
\end{eqnarray}
\begin{theorem}\label{at3}\ \\
 The differential operator which fulfils Eq. \eref{a3.3} is
 \begin{equation}\label{at3.1}
  D_{Nx}^{(\beta)}(B)=\Delta_N^{-1}(x)\det\left[x_a^{N-b}\left(\frac{\partial}{\partial x_a}+(N-b)\frac{\beta}{2}\frac{1}{x_a}+\imath B\right)\right]_{1\leq a,b\leq N}\ .
 \end{equation}
\end{theorem}
\textbf{Derivation:}\\
Kohler \cite{Koh07} has presented a connection between the Jack--polynomials and the matrix Bessel--functions. Let
\begin{equation}\label{a3.4}
 z_a=e^{\imath\frac{2\pi}{L}x_a}\ \ \ {\rm and}\ \ \ n_a=\frac{L}{2\pi}y_a-\left(\frac{N+1}{2}-a\right)\frac{\beta}{2}
\end{equation}
then it is true
\begin{equation}\label{a3.5}
 \varphi_{N}^{(\beta)}\left(\frac{y}{\gamma_2},x\right)=\underset{L\to\infty}{\rm lim}\left(\frac{\Delta_N(z)}{\Delta_N(x)\Delta_N(y)}\right)^{\beta/2}\prod\limits_{a=1}^Nz_a^{-\beta(N-1)/4}J_{N}^{(\beta)}(n,z)\ .
\end{equation}

We expand the determinant in Eq. \eref{a3.1} and have
\begin{eqnarray}\fl
  D_{Nz}(u,\beta)=\nonumber\\
  \fl= \Delta_N^{-1}(z)\sum\limits_{m\in\{0,1\}^N}\prod\limits_{a=1}^N\left[\frac{\beta}{2}\left(z_a\frac{\partial}{\partial z_a}+u\right)\right]^{m_a}\Delta_N(z)\prod\limits_{a=1}^N\left[\left(1-\frac{\beta}{2}\right)\left(z_a\frac{\partial}{\partial z_a}+u\right)\right]^{1-m_a}\hspace*{-1cm}.\label{a3.6}
\end{eqnarray}
Using the substitution \eref{a3.4} and
\begin{equation}\label{a3.7a}
 \widetilde{\Delta}(x)=\prod\limits_{1\leq a<b\leq N}2\imath\sin\left(\frac{\pi}{L}(x_a-x_b)\right)\exp\left(\imath\pi\frac{x_a+x_b}{L}\right)\ ,
\end{equation}
we consider the limit
\begin{eqnarray}
  \fl\underset{L\to\infty}{\lim}\left(\frac{2\pi\imath}{L}\right)^ND_{Nz}(u,\beta)=\nonumber\\
  \fl=  \underset{L\to\infty}{\lim}\frac{1}{\widetilde{\Delta}(x)}\sum\limits_{m\in\{0,1\}^N}\prod\limits_{a=1}^N\left[\frac{\beta}{2}\left(\frac{\partial}{\partial x_a}+\imath\frac{2\pi u}{L}\right)\right]^{m_a}\widetilde{\Delta}(x)\times\nonumber\\
  \fl\times\prod\limits_{j=1}^N\left[\left(1-\frac{\beta}{2}\right)\left(\frac{\partial}{\partial x_a}+\imath\frac{2\pi u}{L}\right)\right]^{1-m_a}=\nonumber\\
  \fl= \Delta_N^{-1}(x)\sum\limits_{m\in\{0,1\}^N}\prod\limits_{a=1}^N\left[\frac{\beta}{2}\left(\frac{\partial}{\partial x_a}+\imath B\right)\right]^{m_a}\Delta_N(x)\left[\left(1-\frac{\beta}{2}\right)\left(\frac{\partial}{\partial x_a}+\imath B\right)\right]^{1-m_a}=\nonumber\\
  \fl= \Delta_N^{-1}(x)\det\left[\frac{\beta}{2}\left(\frac{\partial}{\partial x_a}+\imath B\right)x_a^{N-b}+\left(1-\frac{\beta}{2}\right)x_a^{N-b}\left(\frac{\partial}{\partial x_a}+\imath B\right)\right]_{1\leq a,b\leq N}=\nonumber\\
  \fl= \Delta_N^{-1}(x)\det\left[x_a^{N-b}\left(\frac{\partial}{\partial x_a}+(N-b)\frac{\beta}{2}\frac{1}{x_a}+\imath B\right)\right]_{1\leq a,b\leq N}\ .\label{a3.7}
\end{eqnarray}
Here, we defined a boost $B=\underset{L\to\infty}{\lim}2\pi u/L$ . The eigenvalue in Eq. \eref{a3.2} is in the limit
\begin{equation}\label{a3.8}\fl
 \underset{L\to\infty}{\lim}\left(\frac{2\pi\imath}{L}\right)^N\prod\limits_{a=1}^N\left[n_a+(N-a)\frac{\beta}{2}+u\right]=\prod\limits_{a=1}^N\imath\left(y_a+B\right)={\det}^{1/\gamma_2} \imath(y+ B\eins_{\gamma_2N})\ .
\end{equation}
We assume that Eq. \eref{a3.5} is a uniformly convergent limit. Thus, we combine \eref{a3.5}, \eref{a3.7} and \eref{a3.8} with Eq. \eref{a3.2} and find Eq. \eref{at3.1}.\hfill$\square$

Indeed, $D_{Nx}^{(\beta)}(B)$ is for the unitary case, $\beta=2$,
\begin{equation}\label{a3.10}
 D_{Nx}^{(2)}(B)=\Delta_N^{-1}(x)\prod\limits_{a=1}^N\left(\frac{\partial}{\partial x_a}+\imath B\right)\Delta_N(x)\ .
\end{equation}

\section{Calculation of the supersymmetric Ingham--Siegel integral}\label{app3}

In \ref{app3.1}, we compute the Ingham--Siegel integral. We derive the statements \ref{t1} and \ref{t5} in \ref{app3.2} and \ref{app3.3}, respectively.

\subsection{Decomposition of the Boson--Boson and Fermion--Fermion block integration}\label{app3.1}

We split $\sigma$ in its Boson--Fermion block structure
\begin{equation}\label{4.5}
 \mathfrak{p}\sigma=\left[\begin{array}{cc} \sigma_1 & e^{-\imath\psi/2}\sigma_\eta^\dagger \\ e^{-\imath\psi/2}\sigma_\eta & e^{-\imath\psi}\sigma_2 \end{array}\right]\ .
\end{equation}

The following calculation must be understand in a weak sense. We first integrate over a conveniently integrable function and, then, perform the integral transformations. Hence, we understand $I_k^{(\beta,N)}$ as a distribution where we must fix the underlying set of test--functions. For our purposes, we need Schwartz--functions analytic in the real independent variables.

Since the superdeterminant of $\mathfrak{p}\left(\sigma+\imath\varepsilon\eins_{4k}\right)$ is
\begin{equation}\label{4.6}
 \Sdet\mathfrak{p}\sigma^+=\frac{\det\left(\sigma_1+\imath\varepsilon\eins_{\tilde{k}}\right)}{\det\left[e^{-\imath\psi}\sigma_2+\imath\varepsilon\eins_{\tilde{k}}-e^{-\imath\psi}\sigma_{\eta}\left(\sigma_1+\imath\varepsilon\eins_{\tilde{k}}\right)^{-1}\sigma_{\eta}^\dagger\right]}
\end{equation}
we shift $\sigma_2$ by analytic continuation to $\sigma_2+\sigma_{\eta}\left(\sigma_1+\imath\varepsilon\eins_{\tilde{k}}\right)^{-1}\sigma_{\eta}^\dagger$ and obtain
\begin{eqnarray}\fl
  I_k^{(\beta,N)}(\rho)& = & \int\limits_{\Sigma_{-\psi}^0(\beta,k)}\displaystyle {\rm exp}\left(-\imath\tr r_1\sigma_1+\imath\tr r_2\sigma_2+\imath\tr\left[r_2\sigma_\eta\left(\sigma_1+\imath\varepsilon\eins_{\tilde{k}}\right)^{-1}\sigma_\eta^\dagger\right]\right)\times\nonumber\\
  & \times & \exp\left(\varepsilon\Str r\right) \left[\frac{{\det}\left(e^{-\imath\psi}\sigma_2+\imath\varepsilon\eins_{\tilde{k}}\right)}{{\det}\left(\sigma_1+\imath\varepsilon\eins_{\tilde{k}}\right)}\right]^{N/\gamma_1}d[\sigma]\ .\label{4.7}
\end{eqnarray}
An integration over the Grassmann variables yields
\begin{eqnarray}
  I_k^{(\beta,N)}(\rho) & = & \left(\frac{-\imath\tilde{\gamma}}{2\pi }\right)^{k_1k_2}\exp\left(\varepsilon\Str r\right){\det}^k r_2\times\nonumber\\
  & \times & \int\limits_{\Herm(\beta,k_1)}\exp\left(-\imath\tr r_1\sigma_1\right){\det}\left(\sigma_1+\imath\varepsilon\eins_{\tilde{k}}\right)^{-N/\gamma_1-k}d[\sigma_1]\times\nonumber\\
  & \times & \int\limits_{\Herm(4/\beta,k_2)}\exp\left(\imath\tr r_2\sigma_2\right){\det}\left(e^{-\imath\psi}\sigma_2+\imath\varepsilon\eins_{\tilde{k}}\right)^{N/\gamma_1}d[\sigma_2]\ .\label{4.8}
\end{eqnarray}
With help of Eq. \eref{4.3} we have
\begin{eqnarray}\fl
  I_k^{(\beta,N)}(\rho) & = & \imath^{-k_2N}G_{Nk_1}^{(\beta)} \left(-\frac{\tilde{\gamma}}{2\pi }\right)^{k_1k_2}\displaystyle{\det}^{\kappa} r_1\Theta(r_1)\exp\left(-e^{\imath\psi}\varepsilon\tr r_2\right)\times\nonumber\\
  & \times & {\det}^k r_2\int\limits_{\Herm(4/\beta,k_2)}\exp\left(\imath\tr r_2\sigma_2\right){\det}^{N/\gamma_1}\left(e^{-\imath\psi}\sigma_2+\imath\varepsilon\eins_{\tilde{k}}\right)d[\sigma_2]\ .\label{4.9}
\end{eqnarray}
The remaining integral over the Fermion--Fermion block $\sigma_2$,
\begin{equation}\fl\label{4.10}
 \displaystyle\mathfrak{I}(r_2)=\exp\left(-e^{\imath\psi}\varepsilon\tr r_2\right)\int\limits_{\Herm(4/\beta,k_2)}\exp\left(\imath\tr r_2\sigma_2\right){\det}^{N/\gamma_1}\left(\sigma_2+\imath e^{\imath\psi}\varepsilon\eins_{\tilde{k}}\right)d[\sigma_2]\ ,
\end{equation}
 is up to a constant a differential operator with respect to $r_2$ times the Dirac--distribution of $r_2$ because the determinant term is for $\beta\in\{1,2\}$ a polynomial in $\sigma_2$ and for $\beta=4$ we use Cramers--degeneracy. We give several representations of this distribution.

We first start with an eigenvalue--angle decomposition of $\sigma_2=Us_2U^\dagger$ where $s_2$ is diagonal and $U\in\U^{(4/\beta)}(k_2)$. Integrating over the group $\U^{(4/\beta)}(k_2)$, Eq. \eref{4.10} becomes
\begin{eqnarray}
  \mathfrak{I}(r_2) & = & \displaystyle \exp\left(-e^{\imath\psi}\varepsilon\tr r_2\right)g_{k_2}^{(4/\beta)}\times\nonumber\\
  & \times & \int\limits_{\mathbb{R}^{k_2}}\varphi_{k_2}^{(4/\beta)}(r_2,s_2){\det}^{N/\gamma_1}\left(s_2+\imath e^{\imath\psi}\varepsilon\eins_{\tilde{k}}\right)|\Delta_{k_2}(s_2)|^{4/\beta}d[s_2].\label{4.11}
\end{eqnarray}
For more information about the ordinary matrix Bessel--function
\begin{equation}\label{4.12}
 \varphi_{k_2}^{(4/\beta)}(r_2,s_2)=\int\limits_{\U^{(4/\beta)}(k_2)}\exp\left(\imath\tr r_2Us_2U^\dagger\right)d\mu(U)
\end{equation}
with normalized Haar--measure $d\mu(U)$ see in Ref.  \cite{GuhKoh02,BerEyn08}. The constant $g_{n}^{(\beta)}$ is defined by
\begin{equation}\label{c2}
 \int\limits_{\Herm(\beta,n)}f(H)d[H]=g_{n}^{(\beta)}\int\limits_{\mathbb{R}^{n}}f(E)|\Delta_{n}(E)|^{\beta}d[E]
\end{equation}
independent of a sufficiently integrable function $f$ which is invariant under the action of $\U^{(\beta)}(n)$. The Gaussian distribution is such a function. For the left hand side we obtain
\begin{equation}\label{c3}
 \int\limits_{\Herm(\beta,n)}\exp\left(-\tr H^2\right)d[H]=\gamma_2^{-n(2n-1)/2}2^{-\beta n(n-1)/4}\pi^{n/2+\beta n(n-1)/4}\ .
\end{equation}
The integral on the right hand side is equal to
\begin{equation}\fl\label{c4}
 \int\limits_{\mathbb{R}^{n}}\exp\left(-\gamma_2\sum\limits_{j=1}^nE_j^2\right)|\Delta_{n}(E)|^{\beta}d[E]=\left\{\begin{array}{ll}
                                                              2^{-n(n-5)/4}\prod\limits_{j=1}^{n}\Gamma\left(\frac{j}{2}+1\right) & ,\ \beta=1,\\
							      2^{-n(n-1)/2}\pi^{n/2}\prod\limits_{j=1}^{n}\Gamma\left(j+1\right) & ,\ \beta=2,\\
							      2^{-n(2n-1/2)}\pi^{n/2}\prod\limits_{j=1}^{n}\Gamma\left(2j+1\right) & ,\ \beta=4,
                                                             \end{array}\right.
\end{equation}
see Mehta's book  \cite{Meh67}. Thus, we have
\begin{equation}\label{c5}
 g_{n}^{(\beta)}=\frac{1}{n!}\prod\limits_{j=1}^{n}\frac{\pi^{\beta(j-1)/2}\Gamma\left(\beta/2\right)}{\Gamma\left(\beta j/2\right)}\ .
\end{equation}
This constant is the quotient of the volumes of the permutation group $S(n)$ and of the flag manifold $\U^{(\beta)}(n)/[\U^{(\beta)}(1)]^n$ with the volume element defined as in Ref.  \cite{ZycSom03} denoted by ${\rm Vol}_B$.

We plug the differential operator of \ref{app2} \eref{a3.3} into Eq. \eref{4.11} and have
\begin{eqnarray}
  \mathfrak{I}(r_2) & = & g_{k_2}^{(4/\beta)}\exp\left(-e^{\imath\psi}\varepsilon\tr r_2\right)(\imath\gamma_1)^{-k_2N}\times\nonumber\\
  & \times & \displaystyle  \left[D_{k_2r_2}^{(4/\beta)}\left(\imath e^{\imath\psi}\gamma_1\varepsilon\right)\right]^N\int\limits_{\mathbb{R}^{k_2}}\phi_{k_2}^{(4/\beta)}(r_2,s_2)|\Delta_{k_2}(s_2)|^{4/\beta}d[s_2]\ .\label{4.13}
\end{eqnarray}
The integration over the eigenvalues leads to the Dirac--distribution
\begin{eqnarray}
   \mathfrak{I}(r_2) & = & \displaystyle \left(\frac{2\pi}{\gamma_1}\right)^{k_2}\left(\frac{\pi}{\gamma_1}\right)^{2k_2(k_2-1)/\beta}\frac{\exp\left(-e^{\imath\psi}\varepsilon\tr r_2\right)}{g_{k_2}^{(4/\beta)}}(\imath\gamma_1)^{-k_2}\times\nonumber\\
   & \times & \displaystyle\left[D_{k_2r_2}^{(4/\beta)}\left(\imath e^{\imath\psi}\gamma_1\varepsilon\right)\right]^N\frac{\delta(r_2)}{|\Delta_{k_2}(r_2)|^{4/\beta}}\label{4.14}
\end{eqnarray}
and we find the representation for the supersymmetric Ingham--Siegel integral \eref{4.15}.

\subsection{Derivation of statement \ref{t1}}\label{app3.2}

The boost $\imath e^{\imath\psi}\varepsilon$ in the determinant can simply be shifted away because of
\begin{equation}\fl\label{p1.1}
 D_{k_2r_2}^{(4/\beta)}\left(\imath e^{\imath\psi}\gamma_1\varepsilon\right)\exp\left(\varepsilon e^{\imath\psi}\tr r_2\right)=\hspace*{-0.04cm}\exp\left(\varepsilon e^{\imath\psi}\tr r_2\right)D_{k_2r_2}^{(4/\beta)}(0)=\hspace*{-0.04cm}\exp\left(\varepsilon e^{\imath\psi}\tr r_2\right)D_{k_2r_2}^{(4/\beta)}
\end{equation}
and Eq. \eref{4.14}. Let $\mathfrak{S}$ the set of $\U^{(4/\beta)}(k_2)$--invariant Schwartz--functions on $\Herm(4/\beta,k_2)\rightarrow\mathbb{C}$. The ordinary matrix Bessel--functions are complete and orthogonal in $\mathfrak{S}$ with the sesquilinear scalar product
\begin{equation}\label{p1.2}
 \langle f|f^\prime\rangle=\int\limits_{\mathbb{R}^{k_2}}f^*(x)f^\prime(x)|\Delta_{k_2}(x)|^{4/\beta}d[x]\ .
\end{equation}
The completeness and the orthogonality are
\begin{eqnarray}
  \langle\phi_{k_2}^{(4/\beta)}(x)|\phi_{k_2}^{(4/\beta)}(x^\prime)\rangle & = & \int\limits_{\mathbb{R}^{k_2}}|\phi_{k_2}^{(4/\beta)}(y)\rangle\langle\phi_{k_2}^{(4/\beta)}(y)|\ |\Delta_{k_2}(y)|^{4/\beta}d[y]=\nonumber\\
  & = &  \int\limits_{\mathbb{R}^{k_2}}\phi_{k_2}^{(4/\beta)}(y,x)\phi_{k_2}^{(4/\beta)*}(y,x^\prime)|\Delta_{k_2}(y)|^{4/\beta}d[y]=\nonumber\\
  & = & C_{k}^{(\beta)}\frac{1}{k_2!}\sum_{p\in S(k_2)}\frac{\prod\limits_{j=1}^{k_2}\delta(x_j-x_{p(j)}^\prime)}{|\Delta_{k_2}(x)|^{2/\beta}|\Delta_{k_2}(x^\prime)|^{2/\beta}}\label{p1.3}
\end{eqnarray}
where $S(n)$ is the permutation group of $n$ elements. We defined the constant
\begin{equation}\label{p1.5}
 C_{k}^{(\beta)}=\left(\frac{2\pi}{\gamma_1}\right)^{k_2}\left(\frac{\pi}{\gamma_1}\right)^{2k_2(k_2-1)/\beta}\left(g_{k_2}^{(4/\beta)}\right)^{-2}\ .
\end{equation}
Thus, we write $D_{k_2r_2}^{(4/\beta)}$ in the Bessel--function basis
\begin{eqnarray}\fl
  \qquad\qquad D_{k_2}^{(4/\beta)} & = & {C_{k}^{(\beta)}\ }^{-2}\int\limits_{\mathbb{R}^{k_2}}|\phi_{k_2}^{(4/\beta)}(y)\rangle\langle\phi_{k_2}^{(4/\beta)}(y)|\ |\Delta_{k_2}(y)|^{4/\beta}d[y]\times\nonumber\\
  & \times & D_{k_2x}^{(4/\beta)}\int\limits_{\mathbb{R}^{k_2}}|\phi_{k_2}^{(4/\beta)}(y^\prime)\rangle\langle\phi_{k_2}^{(4/\beta)}(y^\prime)|\ |\Delta_{k_2}(y^\prime)|^{4/\beta}d[y^\prime]=\nonumber\\
  & = & {C_{k}^{(\beta)}\ }^{-1}\int\limits_{\mathbb{R}^{k_2}}{\det}(i\gamma_1y)^{1/\gamma_1}\phi_{k_2}^{(4/\beta)}(y,x)\phi_{k_2}^{(4/\beta)*}(y,x^\prime)|\Delta_{k_2}(y)|^{4/\beta}d[y]\label{p1.6}
\end{eqnarray}
with the action on a function $f\in\mathfrak{S}$
\begin{eqnarray}
  D_{k_2}^{(4/\beta)}|f\rangle & = & {C_{k}^{(\beta)}\ }^{-1}\int\limits_{\mathbb{R}^{k_2}}\int\limits_{\mathbb{R}^{k_2}}{\det}(i\gamma_1y)^{1/\gamma_1}\phi_{k_2}^{(4/\beta)}(y,x)\phi_{k_2}^{(4/\beta)*}(y,x^\prime)f(x^\prime)\times\nonumber\\
 & \times & |\Delta_{k_2}(x^\prime)|^{4/\beta}|\Delta_{k_2}(y)|^{4/\beta}d[x^\prime]d[y]\ .\label{p1.7}
\end{eqnarray}
Due to this representation of the Sekiguchi differential operator analog, $\imath^{k_2}D_{k_2}^{(4/\beta)}$ is symmetric with respect to the scalar product \eref{p1.2}
\begin{equation}\label{p1.8}
 \langle f|\imath^{k_2}D_{k_2}^{(4/\beta)}|f^\prime\rangle=\langle\imath^{k_2}D_{k_2}^{(4/\beta)}f|f^\prime\rangle\ .
\end{equation}

Let $L$ be a real number. Then, we easily see with help of Eq. \eref{at3.1}
\begin{equation}\label{p1.9}
 D_{k_2x}^{(4/\beta)}\det x^{L/\gamma_1}=\prod\limits_{b=1}^{k_2}\left(L+\frac{2}{\beta}b-\frac{2}{\beta}\right)\det x^{(L-1)/\gamma_1}\ .
\end{equation}
Since the property \eref{p1.8}, we obtain for a function $f\in\mathfrak{S}$
\begin{eqnarray}
  & & \int\limits_{\mathbb{R}^{k_2}}\det x^{L/\gamma_1}|\Delta_{k_2}(x)|^{4/\beta}D_{k_2x}^{(4/\beta)}f(x)d[x]=\nonumber\\
  & = & (-1)^{k_2}\int\limits_{\mathbb{R}^{k_2}}f(x)|\Delta_{k_2}(x)|^{4/\beta}D_{k_2x}^{(4/\beta)}\det x^{L/\gamma_1}d[x]=\nonumber\\
  & = & (-1)^{k_2}\prod\limits_{b=1}^{k_2}\left(L+\frac{2}{\beta}b-\frac{2}{\beta}\right)\int\limits_{\mathbb{R}^{k_2}}f(x)|\Delta_{k_2}(x)|^{4/\beta}\det x^{(L-1)/\gamma_1}d[x]\ .\label{p1.10}
\end{eqnarray}
The boundary terms of the partial integration do not appear because $f$ is a Schwartz--function and $D_{k_2x}^{(4/\beta)}$ has the representation \eref{p1.6}.

Let $F$ and $f$ be the functions of statement \ref{t1}. Then, we calculate
\begin{eqnarray}\fl
  \int\limits_{\mathbb{R}^{k_2}}\int\limits_{\Herm(4/\beta,k_2)}F(r_2){\det}^k r_2|\Delta_{k_2}(r_2)|^{4/\beta} \exp\left(\imath\tr r_2\sigma_2\right){\det}^{N/\gamma_1}\left(e^{-\imath\psi}\sigma_2+\imath\varepsilon\eins_{\tilde{k}}\right)d[\sigma_2]d[r_2]=\nonumber\\
  \fl= \int\limits_{\mathbb{R}^{k_2}}\int\limits_{\Herm(4/\beta,k_2)}f(r_2){\det}^{N/\gamma_1} r_2|\Delta_{k_2}(r_2)|^{4/\beta} \exp\left(\imath\tr r_2\sigma_2\right)\times\nonumber\\
  \fl\times{\det}^{N/\gamma_1}\left(e^{-\imath\psi}\sigma_2+\imath\varepsilon\eins_{\tilde{k}}\right)d[\sigma_2]d[r_2]=\nonumber\\
  \fl= \left(\frac{-\imath e^{-\imath\psi}}{\gamma_1}\right)^{k_2N}g_{k_2}^{(4/\beta)}\int\limits_{\mathbb{R}^{k_2}}\int\limits_{\mathbb{R}^{k_2}}f(r_2)\exp\left(\varepsilon e^{\imath\psi}\tr r_2\right)|\Delta_{k_2}(r_2)|^{4/\beta}\times\nonumber\\
  \fl\times {\det}^{N/\gamma_1} s_2|\Delta_{k_2}(s_2)|^{4/\beta}\left(D_{k_2s_2}^{(4/\beta)}\right)^N\phi_{k_2}^{(4/\beta)}(r_2,s_2)d[s_2]d[r_2]=\nonumber\\
  \fl= (\imath e^{-\imath\psi})^{k_2N}g_{k_2}^{(4/\beta)}\prod\limits_{a=1}^N\prod\limits_{b=1}^{k_2}\left(\frac{a}{\gamma_1}+\frac{b-1}{\gamma_2}\right)\times\nonumber\\
  \fl\times \int\limits_{\mathbb{R}^{k_2}}\int\limits_{\mathbb{R}^{k_2}}f(r_2)\exp\left(\varepsilon e^{\imath\psi}\tr r_2\right) |\Delta_{k_2}(r_2)|^{4/\beta}|\Delta_{k_2}(s_2)|^{4/\beta}\phi_{k_2}^{(4/\beta)}(r_2,s_2)d[s_2]d[r_2]=\nonumber\\
  \fl= \left(\frac{2\pi}{\gamma_1}\right)^{k_2}\left(\frac{\pi}{\gamma_1}\right)^{2k_2(k_2-1)/\beta}\frac{\left(\imath e^{-\imath\psi }\right)^{k_2N}}{g_{k_2}^{(4/\beta)}\gamma_1^{k_2N}}\prod_{j=0}^{k_2-1}\frac{\Gamma\left(N+1+2j/\beta\right)}{\Gamma\left(1+2j/\beta\right)}f(0)\ .\label{p1.11}
\end{eqnarray}

The second equality in Eq. \eref{t1.2} is true because of
\begin{equation}\fl\label{p1.12}
 f(0)=\left.\prod\limits_{j=1}^{k_2}\frac{1}{\left(N-k_1\right)!}\left(\frac{\partial}{\partial r_{j2}}\right)^{N-k_1}\left[f(r_2)\exp\left(\varepsilon e^{\imath\psi}\tr r_2\right)\det r_2^{N/\gamma_1-k}\right]\right|_{r_2=0}.
\end{equation}
The function in the bracket is $F$ times the exponential term ${\rm exp}\left(\varepsilon e^{\imath\psi}\tr r_2\right)$.

\subsection{Derivation of statement \ref{t5}}\label{app3.3}

We have to show
\begin{eqnarray}
   & & \int\limits_{\Herm(4/\beta,k_2)}\int\limits_{\Herm(4/\beta,k_2)}F(\rho_2){\det}^k \rho_2\exp\left(\imath\tr \rho_2\sigma_2\right){\det}^{N/\gamma_1}\sigma_2d[\sigma_2]d[\rho_2]\sim\nonumber\\
   & \sim &\int\limits_{\mathbb{R}^{k_2}}F(r_2)\prod\limits_{j=1}^{k}\left(-\frac{\partial}{\partial r_{j2}}\right)^{N-2/\beta}\delta(r_{j2})d[r_2]\label{p5.1}
\end{eqnarray}
for every rotation invariant Schwartz--function $F:\Herm(4/\beta,k_2)\rightarrow\mathbb{C}$ and $\beta\in\{1,2\}$. Due to
\begin{eqnarray}\fl
   \int\limits_{\Herm(4/\beta,k_2)}\hspace*{-0.5cm}\exp\left(\imath\tr r_2\sigma_2\right){\det}\sigma_2^{N/\gamma_1}d[\sigma_2] & \sim & \int\limits_{\mathbb{R}}\int\limits_{\mathbb{R}^{4(k_2-1)/\beta}}\hspace*{-0.5cm}y^N{\rm exp}\left[\imath r_{k_22}\tr(y\eins_{\tilde{\gamma}}+v^\dagger v)\right]d[v]dy\times\nonumber\\
   & \times & \int\limits_{\Herm(4/\beta,k_2-1)}\hspace*{-0.5cm}\exp\left(\imath\tr \tilde{r}_2\tilde{\sigma}_2\right){\det}\tilde{\sigma}_2^{(N+2/\beta)/\gamma_1}d[\tilde{\sigma}_2]\label{p5.2}
\end{eqnarray}
with the decompositions $r_2=\diag\left(\tilde{r}_2,r_{k_22}\eins_{\tilde{\gamma}}\right)$ and
\begin{equation}\label{p5.3}
 \sigma_2=\left[\begin{array}{cc} \tilde{\sigma}_2 & v \\ v^\dagger & y\eins_{\tilde{\gamma}}\end{array}\right]\ ,
\end{equation}
we make a complete induction. Thus, we reduce the derivation to
\begin{equation}\fl\label{p5.4}
   \int\limits_{\mathbb{R}}\int\limits_{\mathbb{R}}\int\limits_{\mathbb{R}^{4(k_2-1)/\beta}}f(x)x^{k_1}y^N{\rm exp}\left[\imath x\tr(y+v^\dagger v)\right]d[v]dydx\sim\int\limits_{\mathbb{R}}f(x)\frac{\partial^{N-2/\beta}}{\partial x^{N-2/\beta}}\delta(x)d[x]
\end{equation}
where $f:\mathbb{R}\rightarrow\mathbb{C}$ is a Schwartz--function. The function
\begin{equation}\label{p5.5}
   \tilde{f}(y)=\int\limits_{\mathbb{R}}f(x)x^{k_1}\exp\left(\imath xy\right)dx
\end{equation}
is also a Schwartz--function. Hence, we compute
\begin{eqnarray}
  & & \int\limits_{\mathbb{R}}\int\limits_{\mathbb{R}}\int\limits_{\mathbb{R}^{4(k_2-1)/\beta}}f(x)x^{k_1}y^N{\rm exp}\left[\imath x\tr(y+v^\dagger v)\right]d[v]dydx=\nonumber\\
  & = & \int\limits_{\mathbb{R}}\int\limits_{\mathbb{R}^{4(k_2-1)/\beta}}\tilde{f}\left[\tr(y+v^\dagger v)\right]y^Nd[v]dy=\nonumber\\
  & = & \int\limits_{\mathbb{R}}\int\limits_{\mathbb{R}^{4(k_2-1)/\beta}}y^{N-2(k_2-1)/\beta}\left(-\frac{\partial}{\partial y}\right)^{2(k_2-1)/\beta}\tilde{f}\left(\tr(y+v^\dagger v)\right)d[v]dy\sim\nonumber\\
  & \sim & \int\limits_{\mathbb{R}}\int\limits_{\mathbb{R}^{+}}\tilde{v}^{2(k_2-1)/\beta-1}\left(-\frac{\partial}{\partial \tilde{v}}\right)^{2(k_2-1)/\beta}\tilde{f}\Bigl(\tr(y+\tilde{v})\Bigr)y^{N-2(k_2-1)/\beta}d\tilde{v}dy\sim\nonumber\\
  & \sim & \int\limits_{\mathbb{R}}\tilde{f}\left(\tr y\right)y^{N-2(k_2-1)/\beta}dy\sim\nonumber\\
  & \sim & \int\limits_{\mathbb{R}}f(x)x^{k_1}\left(-\frac{\partial}{\partial x}\right)^{N-2(k_2-1)/\beta}\delta(x)dx\sim\nonumber\\
  & \sim & \int\limits_{\mathbb{R}}f(x)\frac{\partial^{N-2/\beta}}{\partial x^{N-2/\beta}}\delta(x)d[x]\ ,\label{p5.6}
\end{eqnarray}
which is for $\beta\in\{1,2\}$ well--defined.

\section{Determinantal structure of the $\UOSp(2k/2k)$--Berezinian}\label{app4}

\begin{theorem}\label{at4}\ \\
 Let $k\in\mathbb{N}$, $x_1\in\mathbb{C}^{2k}$ and $x_2\in\mathbb{C}^{k}$. $x_1$ and $x_2$ satisfy the condition
 \begin{equation}\label{a4.1}
  x_{a1}-x_{b2}\neq 0\ \ ,\ \forall a\in\{1,\ldots,2k\}\ \wedge\ b\in\{1,\ldots,k\}\ .
 \end{equation}
 Then, we have
 \begin{equation}\fl\label{a4.2}
  \frac{\Delta_{2k}(x_1)\Delta_{k}^4(x_2)}{V_{k}^2(x_1,x_2)}=(-1)^{k(k-1)/2}\det\left[\left\{\frac{1}{x_{a1}-x_{b2}}\right\}\underset{{1\leq b\leq k}}{\underset{1\leq a\leq 2k}{ }}, \left\{\frac{1}{(x_{a1}-x_{b2})^2}\right\}\underset{{1\leq b\leq k}}{\underset{1\leq a\leq 2k}{ }}\right].
 \end{equation}
\end{theorem}
We prove this theorem by complete induction.\\
\textbf{Derivation:}\\
We rearrange the determinant by exchanging the columns
\begin{eqnarray}
  \det\left[\left\{\frac{1}{x_{a1}-x_{b2}}\right\}\underset{{1\leq b\leq k}}{\underset{1\leq a\leq 2k}{ }}\ ,\ \left\{\frac{1}{(x_{a1}-x_{b2})^2}\right\}\underset{{1\leq b\leq k}}{\underset{1\leq a\leq 2k}{ }}\right]=\nonumber\\
  = (-1)^{k(k-1)/2}\det\left[\frac{1}{x_{a1}-x_{b2}}\ ,\ \frac{1}{(x_{a1}-x_{b2})^2}\right]\underset{{1\leq b\leq k}}{\underset{1\leq a\leq 2k}{ }}\ .\label{a4.3}
\end{eqnarray}
Thus, the minus sign in Eq. \eref{a4.2} cancels out.

We find for $k=1$
\begin{equation}
  \det\left[\begin{array}{cc} \displaystyle\frac{1}{x_{11}-x_{2}} & \displaystyle\frac{1}{(x_{11}-x_{2})^2} \\ \displaystyle\frac{1}{x_{21}-x_{2}} & \displaystyle\frac{1}{(x_{21}-x_{2})^2} \end{array}\right] = \frac{(x_{11}-x_{21})}{(x_{11}-x_{2})^2(x_{21}-x_{2})^2}\ .\label{a4.4}
\end{equation}

We assume that this theorem is for $k-1$ true. Let
\begin{eqnarray}
 s   & = & \left[\frac{1}{x_{a1}-x_{b2}}\ ,\ \frac{1}{(x_{a1}-x_{b2})^2} \right]\underset{{1\leq b\leq k}}{\underset{1\leq a\leq 2k}{ }} = \left[\begin{array}{cc} s_1 & w \\ v & s_2 \end{array}\right]\ , \label{a4.5}\\
 s_1 & = & \left[\begin{array}{cc} \displaystyle\frac{1}{x_{11}-x_{12}} & \displaystyle\frac{1}{(x_{11}-x_{12})^2} \\ \displaystyle\frac{1}{x_{21}-x_{12}} & \displaystyle\frac{1}{(x_{21}-x_{12})^2} \end{array}\right]\ , \label{a4.6}\\
 s_2 & = & \left[\frac{1}{x_{a1}-x_{b2}}\ ,\ \frac{1}{(x_{a1}-x_{b2})^2}\right]\underset{{2\leq b\leq k}}{\underset{3\leq a\leq 2k}{ }}\ , \label{a4.7}\\
 v        & = & \left[\frac{1}{x_{a1}-x_{12}}\ ,\ \frac{1}{(x_{a1}-x_{12})^2}\right]_{3\leq a\leq 2k}\ {\rm and} \label{a4.8}\\
 w        & = & \left[\begin{array}{cc} \displaystyle\frac{1}{x_{11}-x_{b2}} & \displaystyle\frac{1}{(x_{11}-x_{b2})^2} \\ \displaystyle\frac{1}{x_{21}-x_{b2}} & \displaystyle\frac{1}{(x_{21}-x_{b2})^2} \end{array}\right]_{2\leq b\leq k}\ . \label{a4.9}
\end{eqnarray}
Then, we have
\begin{equation}\fl\label{a4.10}
  \det s=\det s_1 \det(s_2-vs_1^{-1}w)\overset{(D.4)}{=}\frac{(x_{11}-x_{21})}{(x_{11}-x_{12})^2(x_{21}-x_{12})^2}\det(s_2-vs_1^{-1}w)\ .
\end{equation}
The matrix in the determinant is equal to
\begin{equation}\fl\label{a4.11}
  (s_2-vs_1^{-1}w)^T=\left[\begin{array}{c} \displaystyle\frac{(x_{11}-x_{a1})(x_{21}-x_{a1})(x_{12}-x_{b2})^2}{(x_{a1}-x_{12})^2(x_{11}-x_{b2})(x_{21}-x_{b2})}\frac{1}{x_{a1}-x_{b2}}\\
  \\ \displaystyle\frac{(x_{11}-x_{a1})(x_{21}-x_{a1})(x_{12}-x_{b2})}{(x_{a1}-x_{12})^2(x_{11}-x_{b2})^2(x_{21}-x_{b2})^2}\frac{P_{ab}}{(x_{a1}-x_{b2})^2} \end{array}\right]\underset{{2\leq b\leq k}}{\underset{3\leq a\leq 2k}{\underset{ }{\underset{ }{\underset{ }{ }}}}}
\end{equation}
where $P_{ab}$ is a polynomial
\begin{eqnarray}\fl
  P_{ab} = (x_{a1}-x_{b2})(x_{11}-x_{b2})(x_{12}-x_{b2})-(x_{a1}-x_{12})(x_{11}-x_{b2})(x_{21}-x_{b2})-\nonumber\\
  \fl-  (x_{21}-x_{b2})(x_{a1}-x_{b2})(x_{11}-x_{12})=\nonumber\\
 \fl= (x_{11}-x_{b2})(x_{21}-x_{b2})(x_{12}-x_{b2})+\nonumber\\
 \fl+ (x_{a1}-x_{b2})\left[(x_{11}+x_{21})(x_{12}+x_{b2})-2x_{11}x_{21}-2x_{12}x_{b2}\right]=\nonumber\\
 \fl= A_b^{(1)}+(x_{a1}-x_{b2})A_b^{(2)}\ .\label{a4.12}
\end{eqnarray}
The polynomials $A_b^{(1)}$ and $A_b^{(2)}$ are independent of the index $a$. Due to the multilinearity and the skew symmetry of the determinant, the result is
\begin{equation}\fl\label{a4.13}
 \det s=\frac{(x_{11}-x_{21})}{(x_{11}-x_{12})^2(x_{21}-x_{12})^2}\frac{\prod\limits_{a=3}^{2k}(x_{11}-x_{a1})(x_{21}-x_{a1})\prod\limits_{b=2}^k(x_{12}-x_{b2})^4}{\prod\limits_{a=3}^{2k}(x_{a1}-x_{12})^2\prod\limits_{b=2}^k(x_{11}-x_{b2})^2(x_{21}-x_{b2})^2}\det s_2
\end{equation}
which completes the induction.\hfill$\square$

\section{Derivation of statement \ref{t0} }\label{app5}

Let $\lambda$ be the wanted eigenvalue and is a commuting variable of the Grassmann algebra constructed from the $\{\tau_q^{(p)},\tau_q^{(p)*}\}_{p,q}$. Then, we split this eigenvalue in its body $\lambda^{(0)}$ and its soul $\lambda^{(1)}$, i.e. $\lambda=\lambda^{(0)}+\lambda^{(1)}$. Let $v$ the $\gamma_2N$--dimensional eigenvector of $H$ such that
\begin{equation}\label{p0.1}
 Hv=\lambda v{\rm\ \ and\ \ }v^\dagger v=1\ .
\end{equation}
In this equation, we recognize in the lowest order of Grassmann variables that $\lambda^{(0)}$ is an eigenvalue of $H^{(0)}$. Then, let $\lambda^{(0)}$ be an eigenvalue of the highest degeneracy $\delta$ of $H^{(0)}$, i.e. $\delta={\rm dim\ ker}(H^{(0)}-\lambda^{(0)}\eins_N)$. Without loss of generality, we assume that $H^{(0)}$ is diagonal and the eigenvalue $\lambda^{(0)}$ only appears in the upper left $\delta\times\delta$--matrix block,
\begin{equation}\label{p0.2}
 H^{(0)}=\left[\begin{array}{cc} \lambda^{(0)}\eins_{\delta} & 0 \\ 0 & \widetilde{H}^{(0)} \end{array}\right]\ .
\end{equation}
We also split the vectors in $\delta$ and $N-\delta$ dimensional vectors
\begin{equation}\label{p0.3}
 v^{(0)}=\left[\begin{array}{c} v_1 \\ v_2 \end{array}\right]{\rm\ \ and\ \ }\tau_q=\left[\begin{array}{c} \tau_{q1} \\ \tau_{q2} \end{array}\right]\ .
\end{equation}
Thus, we find the two equations from \eref{p0.1}
\begin{eqnarray}
 T_{11}v_1-\lambda^{(1)}v_1+T_{12}v_2 & = & 0\label{p0.4}\ ,\\
 T_{21}v_1+\left[\widetilde{H}^{(0)}-\lambda\eins_{N-\delta}+T_{22}\right]v_2 & = & 0\label{p0.5}
\end{eqnarray}
where $T_{nm}=\sum\limits_{q=1}^{\widetilde{N}}l_q\left[\tau_{qn}\tau_{qm}^\dagger+\widetilde{Y}\left(\tau_{qn}^*\tau_{qm}^T\right)\right]$. Eq. \eref{p0.5} yields
\begin{equation}\label{p0.6}
 v_2=-\left[\widetilde{H}^{(0)}-\lambda\eins_{N-\delta}+T_{22}\right]^{-1}T_{21}v_1\ .
\end{equation}
Hence, the body of $v_2$ is zero and we have for Eq. \eref{p0.4}
\begin{equation}\label{p0.7}
 T_{11}v_1-\lambda^{(1)}v_1-T_{12}\left[\widetilde{H}^{(0)}-\lambda\eins_{N-\delta}+T_{22}\right]^{-1}T_{21}v_1 = 0\ .
\end{equation}

If the degeneracy is $\delta>\gamma_2$, we consider a $\delta$--dimensional real vector $w\neq0$ such that $w^\dagger v_1=0$. Then, we get for the lowest order in the Grassmann variables of Eq. \eref{p0.7} times $w^\dagger$
\begin{equation}\label{p0.8}
 w^\dagger T_{11}v_1^{(0)}= 0
\end{equation}
where $v_1^{(0)}$ is the body of $v_1$. The entries of $w^\dagger T_{11}$ are linearly independent. Thus, the body of $v_1$ is also zero. This violates the second property of \eref{p0.1}.

Let the degeneracy $\delta=\gamma_2$. Then, $v_1$ is $\gamma_2$-dimensional and is normalizable. For $\beta=4$, we have the quaternionic case and the matrix before $v_1$ in Eq. \eref{p0.7} is a diagonal quaternion. Hence, it must be true
\begin{equation}\label{p0.9}
 \lambda^{(1)}\eins_{\gamma_2}=T_{11}-T_{12}\left[\widetilde{H}^{(0)}-\lambda\eins_{N-\delta}+T_{22}\right]^{-1}T_{21}\ .
\end{equation}
Considering the second order term in the Grassmann variables of Eq. \eref{p0.9}, $\lambda$'s second order term is $T_{11}$ for $\beta\in\{1,2\}$ and $\tr T_{11}/2$ for $\beta=4$. Eq. \eref{p0.9} is unique solvable by recursive calculation. We plug the right hand side of Eq. \eref{p0.9} into the $\lambda^{(1)}$ on the same side and repeat this procedure. Hence, we define the operator
\begin{eqnarray}
 \fl O(\mu) & = & \frac{1}{\gamma_2}\tr\left\{T_{11}-T_{12}\left[\widetilde{H}^{(0)}-(\lambda^{(0)}+\mu)\eins_{N-\delta}+T_{22}\right]^{-1}T_{21}\right\}{\rm\ and}\label{p0.10}\\
 \fl O^{n+1}(\mu) & = & O\left[O^{n}(\mu)\right]\ .\label{p0.11}
\end{eqnarray}
Then, $\lambda^{(1)}=O^{n}(\lambda^{(1)})$ is true for arbitrary $n\in\mathbb{N}$. The recursion is finished for $n_0\in\mathbb{N}$ if $\lambda^{(1)}=O^{n_0}(\lambda^{(1)})=O^{n_0}(0)$. Due to the Grassmann variables, this recursion procedure eventually terminates  after the $(\gamma_2N\widetilde{N}/2)$'th time. Thus, the eigenvalue $\lambda$ depends on Grassmann variables and is not a real number.

\section*{References}

\end{document}